\documentclass[aps,prd,twocolumn,floats,nofootinbib,longbibliography,showkeys]{revtex4-1}
\usepackage[dvips]{graphicx}

\usepackage{amsmath}
\usepackage{braket}
\usepackage{amsfonts}
\usepackage[colorlinks=true]{hyperref}
\usepackage{hyperref}
\usepackage{xcolor}

\usepackage{booktabs}
\usepackage{aas_macros}

\usepackage{caption}
\usepackage{graphicx}
\usepackage{subcaption}

\usepackage{bm}
\usepackage{slashed}

\usepackage{changes}

\def\aap{\textit{Astronomy and Astrophysics}}

\def\be{\begin{equation}}
\def\ee{\end{equation}}
\def\bea{\begin{eqnarray}}
\def\eea{\end{eqnarray}}

\def\bd{\begin{aligned}}
\def\ed{\end{aligned}}

\def\L{\mathcal{L}}
\def\Lm{\mathcal{L}_m}

\def\t{\tilde}

\newcommand\gammae{\gamma_\textrm{exact}}
\newcommand\betae{\beta_\textrm{exact}}
\newcommand\deltae{\delta_\textrm{exact}}

\definechangesauthor[name={Olivier}, color=magenta]{OM}
\definechangesauthor[name={Aurelien}, color=cyan]{AH}

\begin{document}
\title{Numerical evaluation of the exact post-Newtonian parameters
in Brans-Dicke and entangled relativity theories}

\author{Thomas Chehab}
\email[]{chehab.thomas@oca.eu}

\affiliation{Artemis, Universit\'e C\^ote d'Azur, CNRS, Observatoire C\^ote d'Azur, BP4229, 06304, Nice Cedex 4, France}

\author{Olivier Minazzoli}
\email[]{ominazzoli@gmail.com}

\affiliation{Artemis, Universit\'e C\^ote d'Azur, CNRS, Observatoire C\^ote d'Azur, BP4229, 06304, Nice Cedex 4, France,\\Bureau des Affaires Spatiales, 2 rue du Gabian, 98000  Monaco.}

\begin{abstract}

In the context of Brans-Dicke scalar-tensor theories of gravity, it has recently been shown that the post-Newtonian parameters should be generalized in the context of strongly gravitating bodies, and that its generalization---the so-called \textit{exact parameters}---actually depends on the pressure and energy density of the considered celestial body. Here we develop two new methods to numerically obtain the \textit{exact parameters} by means of the usual Tolman-Oppenheimer-Volkoff computation, and find that the difference from the value of standard post-Newtonian parameters can be more than 80\% in some situations. We also provide the connection with the Damour-Esposito-Farèse non-perturbative parameter $\alpha_{DEF}$. We then apply the methodology to the case of Entangled Relativity, and derive these exact parameters for the Sun and the Earth, as well as for neutron stars. We argue that current and foreseeable experiments are likely able to constrain the theory under the assumption that $\Lm=-\rho$, where $\rho$ is the total energy density. If $\Lm=T$ instead, as often advocated in the literature, then there is no deviation with respect to General Relativity and the prospects of testing Entangled Relativity become much more remote in time, as only compact objects with extreme electric or magnetic fields could lead to some deviation from General Relativity. 
\end{abstract}

\maketitle

\section{Introduction}
\label{sec:intro}
One of the first alternatives to General Relativity was the Brans–Dicke theory of gravity. The main idea behind this theory is the hypothesis that gravity is mediated not only through the metric tensor $g_{\mu \nu}$, but also by a scalar field $\phi$. These theories are known as scalar-tensor theories of gravity \citep{brans_Dicke:1961pr}. In such frameworks, the scalar field is inversely proportional to the Newtonian gravitational constant $G$. Later, the Bergmann-Wagoner-Nordtvedt theory generalizes the Brans-Dicke theory of gravity using the assumption that the Brans-Dicke coupling parameter $\omega$ can itself depend on the scalar field of the considered scalar-tensor theory \cite{Bergmann1968, wagoner:1970, nordvert:1968}.

Since the first formulation of General Relativity, we have sought to test the theory experimentally. First, through observations of Mercury's perihelion, the deflection of light by the Sun, and gravitational redshift. These experiments were successful and confirmed the predictions of General Relativity, thereby lending strong support to the theory. Later, the Shapiro delay effect \cite{shapiro:1964prl} was verified experimentally, and nowadays new constraints are obtained from the orbital decay of binary pulsars \cite{taylor:1982}, gravitational waves \cite{Ligo}, direct observations of the shadows of supermassive black holes \cite{Akiyama_2019}, and highly precise tests of the strong equivalence principle \cite{williams:2009ij}---see \cite{will:2014lr,abbott:2020lr,fienga:2024lr,EHT:2025lr} for reviews. Among all these tests, the usual Solar-System constraints are derived from post-Newtonian studies. In the weak-field regime, one can expand the metric in powers of $v/c \approx \sqrt{GM/rc^2}$, which leads to coefficients called post-Newtonian parameters. These parameters quantify deviations from General Relativity (for a review of experimental tests of General Relativity, see \cite{Will_2014}).

Recently, in the context of Brans–Dicke theory, using the Brans–Dicke class I solution \citep{brans_Dicke:1961pr}, a novel \textit{`exact'} expression of the post-Newtonian parameters $\beta$ and $\gamma$ has been derived \citep{chauvineau:2024plb} and later in \cite{chauvineau:2024prd}, these parameters (plus $\delta$, a higher-order parameter) have been derived in Bergmann-Wagoner-Nordvert theory. They have been shown to depend explicitly on the energy density and pressure of the considered body and are \textit{exact} and \textit{non-perturbative} quantities, but have so far never been evaluated numerically. Indeed, because these new exact quantities directly depend on the structure of the considered bodies, one needs to integrate the interior solution numerically---in contrast to the case of usual (\textit{perturbative}) post-Newtonian parameters that can be derived analytically.

In this communication we provide a numerical evaluation of these exact (non-perturbative) post-Newtonian parameters in the context of the Brans-Dicke theory and then in Entangled Relativity. In order to compute their numerical values we will use the Tolman-Oppenheimer-Volkoff framework following \cite{arruga:2021pr,arruga:2021ep,chehab:2025cq}.\\

This paper is organized as follows. In Sec.~\ref{sec:BD}, we review the notion of exact post-Newtonian parameters in the framework of Bergmann--Wagoner--Nordtvedt scalar--tensor theories and summarize the analytical expressions previously derived for spherical bodies, emphasizing their dependence on the internal structure of the source. In Sec.~\ref{sec:TOV}, we derive the Tolman--Oppenheimer--Volkoff equations in Brans--Dicke theory and explain how these equations can be used to numerically evaluate the exact post-Newtonian parameters. In Sec.~\ref{sec:gammaBD}, we present a detailed numerical study of these parameters in Brans--Dicke gravity, including a cross-check based on external analytical solutions. In addition, we provide the connection with Damour-Esposito-Farèse non-perturbative parameter $\alpha_{DEF}$. In Sec.~\ref{sec:ER}, we introduce Entangled Relativity, discuss its usual post-Newtonian limit, and derive the exterior solution necessary to compute its exact post-Newtonian parameters. Sec.~\ref{sec:gammaER} is devoted to the numerical evaluation of these parameters in Entangled Relativity, first in the solar system and then for compact objects such as neutron stars. Sec.~\ref{sec:GW} infers the level of dipolar gravitational waves emission in Entangled Relativity from the exact parameters, for a binary system composed of a pulsar and a white dwarf. Finally, Sec.~\ref{sec:concl} summarizes our main results.

\section{Exact Post-Newtonian parameters in Bergmann-Wagoner-Nordtvedt theory}

\label{sec:BD}

Here we well briefly review the results obtain by \cite{chauvineau:2024prd}, in which is derived a generalisation of the exact post-Newtonian parameters in Bergmann-Wagoner-Nordtvedt theory. Its action reads
\small
\be
\label{eq:action_BWN}
S = \frac{1}{16\pi}\int d^{4}x \sqrt{-g} \left[  \Phi R   - \frac{\omega(\Phi)}{\Phi} \nabla_{\mu}\Phi \nabla^{\mu}\Phi \right] + S_{\rm m}[g_{\mu\nu},\Psi] ,
\ee
\normalsize
where $\Phi$ is a scalar field, $\omega(\Phi)$ a coupling function that depends on the scalar field and $S_m$ the matter action, which depends on the metric $g$ and the matter fields $\Psi$. Using the least action principle one can obtain the following field equations

\be
\begin{aligned}\label{FE_BWN}
 G_{\mu\nu} =& \frac{8\pi}{\Phi} T_{\mu\nu}
+ \frac{1}{\Phi} \left[ \nabla_{\mu}\nabla_{\nu}
- g_{\mu\nu}\Box \right] \Phi
\\ &+ \frac{\omega(\Phi)}{\Phi^2}
\left(  \nabla_{\mu}\Phi\nabla_{\nu}\Phi
  - \tfrac{1}{2} g_{\mu\nu} (\nabla\Phi)^2
\right) ,
\end{aligned}
\ee
\begin{equation}\label{FE_Phi}
\Box\Phi = \frac{1}{2\omega(\Phi)+3}
\left[  8\pi T  - \frac{d\omega}{d\Phi} \, (\nabla\Phi)^2\right] .
\end{equation}
\subsection{Summary of previous results in general scalar-tensor theory}
\label{sec:sumChauvineau}

To derive the formulation of the exact post-Newtonian parameters, the author of \cite{chauvineau:2024prd} started in Einstein frame with the Janis-Newman-Winicour metric \citep{janis:1968prl} in isotropic coordinates that they re-expressed in the Jordan frame thanks to a conformal transformation. This latter reads

\be \label{eq:CT}
d\varphi = \sqrt{3+2\omega (\Phi)} d\ln{\Phi}.
\ee
While the metric solution reads
\be
\label{eq:JNW_JF}
ds^{2} = -A dt^{2} + B dr^{2} + C d\Omega^{2},
\ee
Where the functions A, B and C are given by 
\be\label{eq:BC1}
\begin{cases}
A = \frac{1}{\Phi}\left( \frac{r-k}{r+k} \right)^{2d}, \\
B = \frac{1}{\Phi}\left( 1 + \frac{k}{r} \right)^{4} 
    \left( \frac{r-k}{r+k} \right)^{2 - 2d}, \\
C = B r^2.
\end{cases}
\ee
In which $k$ is linked to the mass by $k = \frac{2m}{d}$ and $d$ is the scalar charge. 
They assumed a static and spherically symmetric stress energy tensor defined by 
\be
\left( T^{b}_{a} \right) = \operatorname{diag}(-\rho, \, p_{\parallel}, \, p_{\perp}, \, p_{\perp}).
\ee
With $\rho$ the energy density and $p_{\parallel}$, $p_{\perp}$ are respectively the parallel and perpendicular pressure. Both are functions of $r$. By focusing on the (00) component of Eq. (\ref{FE_BWN}) and the scalar field equation Eq. (\ref{FE_Phi}) one can derive the value of $d$ and $k$ as
\begin{equation}
\left\{
\begin{aligned}
\label{eq:d}
d &= \frac{E^* + P^*}{\sqrt{(E^* + P^*)^2 + (E^*_{\omega} - P^*_{\omega})^2}},\\
k &= \tfrac{1}{2} \sqrt{(E^* + P^*)^2 + (E^*_{\omega} - P^*_{\omega})^2},
\end{aligned}
\right.
\end{equation}
with
\be\label{eq:integrals_BD}
\begin{cases} 
E^{*} = \int_{(star)} \left(-T^0_0\right) dV,\\
P^{*} = \int_{(star)}  T^i_i dV,\\
E^{*}_\omega = \int_{(star)} \frac{(-T^0_0)}{\sqrt{3+2\omega(\Phi)}}  dV,\\
P^{*}_\omega = \int_{(star)} \frac{(T^i_i)}{\sqrt{3+2\omega(\Phi)}}  dV,\\
\end{cases}
\ee
where $E^{*}$, and $P^*$ are respectively the total integrated energy and total integrated pressure while $E^{*}_\omega$, and $P^*_\omega$ are the total integrated energy and pressure divided by a function of $\omega(\Phi)$. Here $dV = 4\pi \sqrt{AB}C dr$. We see with Eq. (\ref{eq:d}) that the scalar charge $d$ explicitly depend on the structure of the considered body.

By Taylor expanding $A$ and $B$ one can obtain

\begin{equation}
\left\{
\begin{aligned}
\label{eq:taylor_A_B}
A &= 1 + x \left[ \frac{d}{dx} \left( A \right) \right]_0 
   + \tfrac{1}{2} x^2 \left[ \frac{d^2}{dx^2} \left( A \right) \right]_0 
   + O(x^3),\\
B &= 1 + x \left[ \frac{d}{dx} \left( B \right) \right]_0 
   + \tfrac{1}{2} x^2 \left[ \frac{d^2}{dx^2} \left( B \right) \right]_0 
   + O(x^3),
\end{aligned}
\right.
\end{equation}
with $x$ given by $x = \frac{k}{r}$. The subscript $0$ indicates that the value is taken at $x=0$ (i.e. : $r \rightarrow \infty$). By solving the second and third term of $A$ and $B$ one can compute the total metric expansion and thus identify, by comparing with usual post-Newtonian definition, the mass and the exact post-Newtonian parameters as

\begin{subequations}\label{eq:exact_param}
\begin{align}
\label{eq:mass_bd}
m 
&= 2\left( d + \sqrt{\frac{1 - d^2}{3+2\omega_0}} \right) k, \\
\beta_{\text{exact}} 
&= 1 + \left( \frac{ \left( \frac{d\omega}{d\Phi} \right)_0 }{ \left(3+2\omega_0\right)^2 } \right) 
\frac{1 - d^2}{\left( d + \sqrt{ \frac{1 - d^2}{3+2\omega_0} } \right)^2}, \\
\label{eq:gammae}
\gamma_{\text{exact}} 
&= \frac{ d - \sqrt{ \frac{1 - d^2}{3+2\omega_0} } }{ d + \sqrt{ \frac{1 - d^2}{3+2\omega_0} } }, \\
\label{eq:deltae}
\delta_{\text{exact}}
&= \begin{aligned}[t]
   \frac{4}{3} & \left( \beta_{\text{exact}} + \gamma_{\text{exact}}^2 - 1  \right) -\\
   & \frac{1}{3} \left( d + \sqrt{ \frac{1 - d^2}{3+2\omega_0} } \right)^{-2},
   \end{aligned}
\end{align}
\end{subequations}
where $\omega_0$ is the background value of $\omega(\Phi)$.
As a reminder, $d$ is a parameter of the Janis-Newman-Winicour metric (Eq. ({\ref{eq:JNW_JF})) and is given by Eq. (\ref{eq:d}). It explicitly depends on the structure of the star. Hence the post-Newtonian parameters $\beta_{exact}, \gamma_{exact}, \delta_{exact}$ are formulated with an explicit dependence on the structure of the considered body. This result was obtained using isotropic Janis-Newman-Winicour metric. Using this result one can replace d with Eq. (\ref{eq:d}) an re-arrange the equation to obtain

\begin{subequations} \label{eq:exact_param_theta}
\bea
&&m = \left( 1 + \frac{\Xi}{\sqrt{3 + 2\omega_0}} \right)(E^{*} + P^{*}) , \label{eq:1a} \\
&&\frac{\sigma}{m} = \frac{\Xi}{\sqrt{3 + 2\omega_0} + \Xi} , \label{eq:1b}\\
&&\betae = 1 + \frac{\left( \frac{d\omega}{d\Phi} \right)_0}{3 + 2\omega_0} 
           \left( \frac{\sigma}{m} \right)^2 , \label{eq:1c}\\
&&\gammae = \frac{\sqrt{3 + 2\omega_0} - \Xi}{\sqrt{3 + 2\omega_0} + \Xi} , \label{eq:1d} \\
&& \deltae = \frac{4}{3} \left[ \beta + \gamma^{2} - 1 
           - \frac{(3 + 2\omega_0)(1 + \Xi^{2})}
                  {4\left(\sqrt{3 + 2\omega_0} + \Xi \right)^{2}} \right] , \label{eq:1e}
\eea
\end{subequations}
in which $\Xi = \frac{1}{\sqrt{3+2\omega_0}} \frac{1-\Theta}{1+\Theta}$ with $\Theta = \frac{P^*}{E^*}$. Detailed calculation are presented in \cite{chauvineau:2024prd}.

\subsection{Connection with observable quantities}
\label{sec:obs}

Let us stress that the exact post-Newtonian parameters characterize the exact external metric of compact objects and are therefore directly related to observable quantities, since the trajectories of particles and light surrounding the compact object depend explicitly on these parameters. This can be understood simply by noting that they are obtained by Taylor expanding the exact external solution of compact bodies, given by Eqs. (\ref{eq:JNW_JF}-\ref{eq:BC1}).

In particular, once the exact parameters are known, the Shapiro delay and light deflection follow directly from \cite{linet:2013cq,linet:2016pr} through the Time Transfer Function formalism.\footnote{Let us note, however, that $\mathcal{B}$ in the notation of \cite{linet:2013cq} corresponds to $B^{-1}$ in ours.}

Their relation to pulsar timing observations will be further investigated in Secs. \ref{sec:DEFvsGammae} and \ref{sec:GW}.

\subsection{Invariance under radial coordinate transformation}
\label{sec:invR}

An important point lies in the fact that the integrals presented in equations (\ref{eq:integrals_BD}) are invariant under a radial coordinate change. Indeed if one performs such a change $(t,r,\theta, \phi) \rightarrow (t' = t,r'(r),\theta' = \theta, \phi' = \phi)$ in the end, the volume element gives : $\sqrt{A'B'}C'dr' = \sqrt{AB\left(\frac{dr}{dr'}\right)^2}C dr'$. And hence $dV = dV' \text{sign}\left(\frac{dr}{dr'} \right) $. Then, the sign of $\frac{dr}{dr'} $ can be reabsorbed into the  permutation of lower and upper bound in the integral. Lastly, $T^0_0$ and $T^i_i$ are scalars. Hence these integrals are invariant through a radial coordinate change.

This is important because the Tolman-Oppeheimer-Volkoff metric does not use the same radial coordinate as \cite{chauvineau:2024prd}. Hence, this invariance is a necessary property in order to be able to inject the Tolman-Oppeheimer-Volkoff numerical evaluations of the various quantities in the analytical expressions provided in \cite{chauvineau:2024prd}---summarized in Sec. \ref{sec:sumChauvineau}.
\subsection{Difference between exact and usual post-Newtonian parameters}

It is worth emphasizing that the \textit{exact} post-Newtonian parameters in equations (\ref{eq:exact_param}-\ref{eq:exact_param_theta}) depend on the internal structure of each body, whereas the usual post-Newtonian parameters in the literature, see e.g. \cite{will:2014lr,fienga:2024lr}, only depend on the specific theory considered. 

The difference between the two sets of post-Newtonian parameters is that the usual post-Newtonian parameters are derived in the weak-field limit of a given theory---in particular, assuming that $\rho\gg P$---whereas the \textit{exact post-Newtonian parameters} are \textit{non-perturbative} quantities that are also valid for ultra-relativistic compact bodies---for which $P\sim \rho$. 

They are \textit{exact} in the sense that they correspond to the Taylor expansion of the exact exterior metric of a compact body, see Eqs. (\ref{eq:JNW_JF}-\ref{eq:BC1}), whereas standard post-Newtonian parameters are derived by assuming that the stress-energy tensor satisfies the usual perturbative post-Newtonian condition on matter fields, namely $T^{\alpha \beta}=\mathcal{O}(c^2,c^1,c^0)$---see \cite{fienga:2024lr} and references therein. Therefore, exact post-Newtonian parameters generalize the standard post-Newtonian parameters to non-perturbative regimes.

However, whereas standard post-Newtonian parameters are independent of the internal structure of bodies and depend solely on the theory under consideration, exact post-Newtonian parameters depend explicitly on the internal structure. Indeed, the issue with the exact parameters in equations (\ref{eq:exact_param}-\ref{eq:exact_param_theta}) is that one needs to know the internal structure of the star---through $\rho, \, p_{\parallel}$ and $p_{\perp}$---as well as the internal metric---through $A$ and $B$---in order to evaluate them.

The independence of standard post-Newtonian parameters from the internal structure of bodies only comes from the approximate assumption that leads to their derivation, namely that $T^{\alpha \beta}=\mathcal{O}(c^2,c^1,c^0)$. In particular, in the limit where this assumption becomes valid---as in the Solar System \cite{fienga:2024lr}, for instance---the exact post-Newtonian parameters reduce to the standard ones, and their independence from the internal structure becomes manifest within this approximation---as one can check from Eqs. (\ref{eq:exact_param_theta}).\\

In what follows, we show how to evaluate the exact post-Newtonian parameters numerically by solving the Tolman-Oppeheimer-Volkoff equations.

\section{Tolman-Oppenheimer-Volkoff equations in Brans-Dicke theory - Rewriting the exact post-Newtonian parameters}
\label{sec:TOV}

In the Bergmann-Wagoner-Nordtvedt theory, the parameter $\omega$ is a function of $\phi$. As a consequence, the Bergmann-Wagoner-Nordtvedt theory actually corresponds to an infinite class of theories---one for each function $\omega(\Phi)$ considered---and one needs to restrict oneself to a specific theory in order to derive the corresponding phenomenology. We will focus on the particular case $\omega(\phi) = \omega$ which corresponds to Brans-Dicke theory. Current constraints on the Strong Equivalence Principle with a pulsar in a triple star system give $\omega > 140~000$ \cite{voisin:2020aa}.

We first derive the modified Tolman-Oppenheimer-Volkoff equations of the Brans-Dicke theory that we will solve numerically in order to evaluate the exact post-Newtonian parameters. Then, with numerical integration we will directly evaluate them. 

\subsection{Tolman-Oppenheimer-Volkoff equations of the Brans-Dicke theory}

To derive the Tolman-Oppenheimer-Volkoff equations in Brans-Dicke theory, we will follow the same method as in \cite{arruga:2021pr}. The Brans-Dicke theory \citep{brans_Dicke:1961pr} is defined by the action 

\be
\label{eq:SBD}
S_{BD} = \int d^4_gx \left[\frac{1}{2 \Tilde \kappa} \left( \phi R - \frac{\omega}{\phi} \partial_\sigma \phi \partial^\sigma \phi \right) + \Lm \right].
\ee

Using the least action principle one can obtain the following field equations

\be
\begin{aligned}
G_{\mu \nu} = \frac{\Tilde \kappa}{\phi} T_{\mu \nu} +&\frac{\omega}{\phi^2} \left( \partial_\mu \phi \partial_\nu \phi - \frac{1}{2} g_{\mu \nu} \partial_\sigma \partial^\sigma \phi \right) \\+ &\frac{1}{\phi} \left( \nabla_\mu \nabla_\nu \phi - g_{\mu \nu} \Box \phi \right),
\label{eq:BD_FE} 
\end{aligned}
\ee

\be
\label{eq:boxphi}
\Box\phi = \frac{\t \kappa}{3+2\omega} T,
\ee
with $\Tilde \kappa = \frac{8 \pi G}{c^4}$, with $G$ the Newton constant. We now set : 

\be
D^\alpha_\beta = \frac{1}{\phi}  \left( \nabla^\alpha \nabla_\beta - \delta^\alpha_\beta \Box \right)\phi,
\ee
and

\be
H^\alpha_\beta = \frac{\omega}{\phi^2} \left( \partial^\alpha \phi \partial_\beta \phi - \frac{1}{2} \delta^\alpha_\beta \partial_\sigma \partial^\sigma \phi \right).
\ee

Using the (00) component of the field equation (\ref{eq:BD_FE}), we obtain an equation over $g_{rr} \equiv b$.

\be
\frac{\dot{b}}{b} = rb \left( \frac{\t \kappa \rho}{\phi} - H^0_0 - D^0_0 \right) + \frac{1-b}{r}.
\label{eq:bdotb}
\ee
While one finds : 
\be
D^0_0 = \frac{\dot{a}}{2ab} \frac{\dot\phi}{\phi} + \frac{\t \kappa}{3+2\omega} \frac{T}{\phi},
\ee
and
\be
H^0_0 = - \frac{\omega}{2b}\frac{\dot\phi^2}{\phi^2},
\ee
in which $a \equiv g_{tt} $.
With a direct integration of Eq. (\ref{eq:bdotb}), one obtains 

\be
b = \frac{1}{1-\frac{m \t \kappa }{4 \pi r}},
\ee

setting 

\be
m = \frac{4\pi}{\t \kappa} \int_0^r dr' r'^2x \left( \frac{\t \kappa \rho}{\phi} - H^0_0  - D^0_0 \right).
\ee
Thanks to which we can derive a differential equation for the mass 
\be
\Dot m = \left( \frac{\t \kappa \rho}{\phi} - H^0_0 - D^0_0 \right) \frac{4 \pi r^2}{\t \kappa} .
\ee
Proceeding similarly with the (11) component of the field equation (\ref{eq:BD_FE}), leads to differential equation for $a$
\be
\begin{aligned}
\frac{\dot{a}}{a} = \frac{b}{r} &\left( \frac{r^2 \t \kappa P }{\phi} + \frac{\dot\phi^2}{\phi^2} \frac{r^2 \omega}{2b} + 1 \right) \left( 1+\frac{r}{2}  \frac{\dot\phi}{\phi}  \right)^{-1} \\ & -\frac{b}{r}\left( \frac{1}{b} + \frac{2 r}{b} \frac{\dot\phi}{\phi}  \right) \left( 1+\frac{r}{2}  \frac{\dot\phi}{\phi}  \right)^{-1}.
\end{aligned}
\ee
Concerning the pressure, using the conservation equation of the theory $ \nabla_\sigma T^{\alpha \sigma} = 0$, one can obtain a differential equation representing its behaviour
\be\label{eq:dpdr_BD}
\dot{P} = - \left( \rho +P \right) \frac{\dot{a}}{2a}.
\ee
Lastly, with the scalar field equation (\ref{eq:boxphi}), one obtains a differential equation of the scalar field 
\be
    \Ddot{\phi} = -\frac{\Dot{\phi}}{2} \left( \frac{\dot{a}}{a} - \frac{\dot{b}}{b} + \frac{4}{r}\right) + b\Box \phi.
\ee
\paragraph{\bf{Summary of equations :}}

Up to now, every necessary quantity has been expressed as a differential equation that can be integrated numerically to model compact objects in Brans–Dicke theory. 

\be \label{eq:summary_BD}
\left\{
\begin{aligned}
    \dot{P} &= - \left( \rho + P \right)\frac{\dot{a}}{2a}, \\[6pt]
    \Dot m &= \left( \frac{\t \kappa \rho}{\phi} - H^0_0 - D^0_0 \right)\frac{4\pi r^2}{\t \kappa}, \\[6pt]
    \Dot \phi &= \psi, \\[6pt]
    \dot{\psi} &= -\frac{\psi}{2}\left( \frac{\dot{a}}{a} - \frac{\dot{b}}{b} + \frac{4}{r} \right) + b \Box \phi.
\end{aligned}
\right.
\ee

\subsection{The equation of state}

The goal of this study is to numerically evaluate exact post-Newtonian parameters. Hence our goal is only to obtain their broad behaviour. As a first exploratory model, we decided to use a simple polytropic equation of state defined by 

\be \label{eq:eos}
P=K \rho^{\gamma},
\ee
with $\gamma = 5/3$, is the adiabatic index and $K = 1.475 \times 10^{-3}$ (fm$^3$/MeV)$^{2/3}$ is simply a constant \citep{arbanil:2013pr,arruga:2021pr}. Moreover, we neglect magnetic field and rotational effects, for the reason explained above. 

\subsection{Solving the Boundary Value Problem}

Solving these equations is a Boundary Value Problem. Indeed, one wants to find the initial condition at the center of the star that will lead to the desired normalization at infinity for any value of the central density. To resolve this problem we used a normalization technique \cite{arruga:2021pr}. Indeed one can remark that if ($\phi$, $\psi$, $m$, $P$)($r$) denotes a solution of equation (\ref{eq:summary_BD}), then another solution is given by ($\alpha \phi$, $\alpha^{5/4} \psi$, $\alpha^{-1/4} m$, $\alpha^{3/2} P$)($\alpha^{-1/4} r$) for any $\alpha$. Then if a simulation starts with an initial value $\phi_0$ that leads to a given star radius $\mathcal{R}_*$ with an asymptotical scalar field value $\phi_\infty$, then the radius with the appropriate unit is given by $R_* = \phi_\infty^{1/4} \mathcal{R}_*$, when the property has been used with $\alpha = \phi_\infty$, such that the final normalization of $\phi$ at infinity is one.


\subsection{Exact post-Newtonian parameter in Brans-Dicke gravity}
\label{sec:exactPNBD}

As explained in sec. \ref{sec:invR}, the equations for the exact post-Newtonian parameters given in Eq. (\ref{eq:exact_param_theta}) are invariant under radial coordinate change. Hence even if they have been derived in \cite{chauvineau:2024prd} using Janis-Newman-Winicour metric in isotropic coordinate, the result is the same in a Tolman-Oppenheimer-Volkoff spherical metric. 

In the framework of Brans-Dicke theory, as $\omega = cst$ one can rewrite Eq. (\ref{eq:exact_param_theta})  as

\begin{subequations}\label{eq:exact_param_theta_BD}
\bea
\betae &&= 1, \\
\gammae&&= \frac{1+\omega+\Theta(2+\omega)}{2+\omega+\Theta(1+\omega)}, \\
\deltae&&= \frac{4}{3} \left[ \gammae^2
          - \frac{(3+2\omega) ( 1 + \Xi^2)}
          {4 \left( \sqrt{3+2\omega} + \Xi \right)^2} \right].
\eea
\end{subequations}
With $\Xi = \frac{1}{\sqrt{3+2\omega}} \left( \frac{1-\Theta}{1+\Theta} \right)$ in Brans-Dicke theory. As a reminder $\Theta = \frac{P^*}{E^*}$, with 
\be \label{eq:tointegrate}
X^* = \int_0^{\rho*} 4\pi X \sqrt{ab^3} \rho^2 d\rho,
\ee
where $\rho$ is the radial coordinate in Tolman-Oppenheimer-Volkoff framework. Then, numerical evaluations of $a$, $b$, $\rho$ and $P$ allows one to evaluate Eq. (\ref{eq:tointegrate}) numerically, from which one infers the values of the exact post-Newtonian parameters in Eq. (\ref{eq:exact_param_theta_BD}).

\section{Numerical evaluation of exact parameters in Brans-Dicke theory}
\label{sec:gammaBD}


\subsection{Numerical estimation of $(\gammae, \betae, \deltae)$ for neutron stars}

We can now evaluate exact post-Newtonian parameters numerically. Indeed using numerical integration, we solve the Tolman-Oppenheimer-Volkoff equations in the framework of the Brans-Dicke theory.

In figures \ref{plot:edd_m} and \ref{plot:relative} we we plotted color maps that represent  $1 - \gammae$ and $\frac{\gammae - \gamma_{PN}}{1 - \gamma_{PN}}\%$, respectively, to probe their respective deviations for various values of the parameter $\omega$ and the central energy density $\rho_c$. In figures \ref{plot:deltae} and \ref{plot:delta_relative}, the shade of colour represents respectively $1 - \deltae$ and $\frac{\deltae - \delta_{PN}}{1 - \delta_{PN}}\%$. Here $\gamma_{PN}$ and $\delta_{PN}$ are the standard (perturbative) formulation of the post-Newtonian parameters $\gamma$ and $\delta$ and are given by $\gamma_{PN} = \frac{1+\omega}{2+ \omega}$ and $\delta_{PN} = \frac{4}{3} \left( \gamma_{PN}^2 + \beta_{PN} \right) - \frac{\gamma}{6} - \frac{3}{2}$ \cite{T_Damour_1992,Minazzoli_CQG_2012}.


One can notice in all figures, that for a same value of $\omega$, the resulting color values explicitly vary with respect to the  density of the star, which is expected as the exact post-Newtonian parameters depend on the structure of the object studied.

\begin{figure}[h]
	\begin{center}
		\includegraphics[scale=0.45]{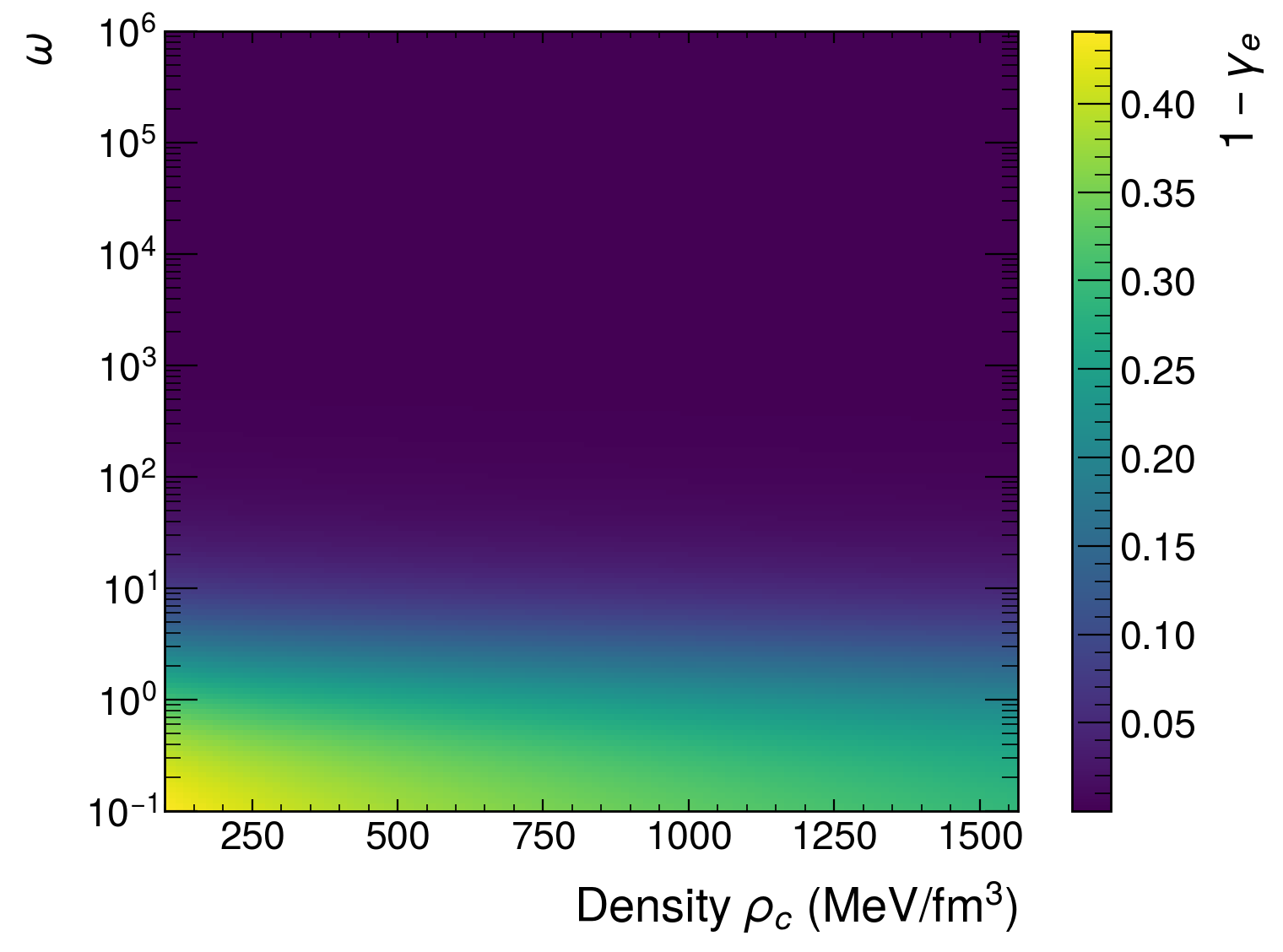}
\caption{Plot of $1-\gammae$ for 150 different values of $\omega$ and of central densities. The values of $\omega$ ranges from $10^{-1}$ to $10^6$ meanwhile the values of central density ranges from $100$ MeV/fm$^3$ to 1570 MeV/fm$^3$.  This limit comes from the fact that we want the speed of sound in the neutron star to be less that the conservative limit considered of $c/\sqrt 3$  \citep{bedaque:2015pr}. The deviation of $\gammae$ from unity is represented in colour shades.} \label{plot:edd_m}
	\end{center}
\end{figure}

\begin{figure}[h]
	\begin{center}
		\includegraphics[scale=0.45]{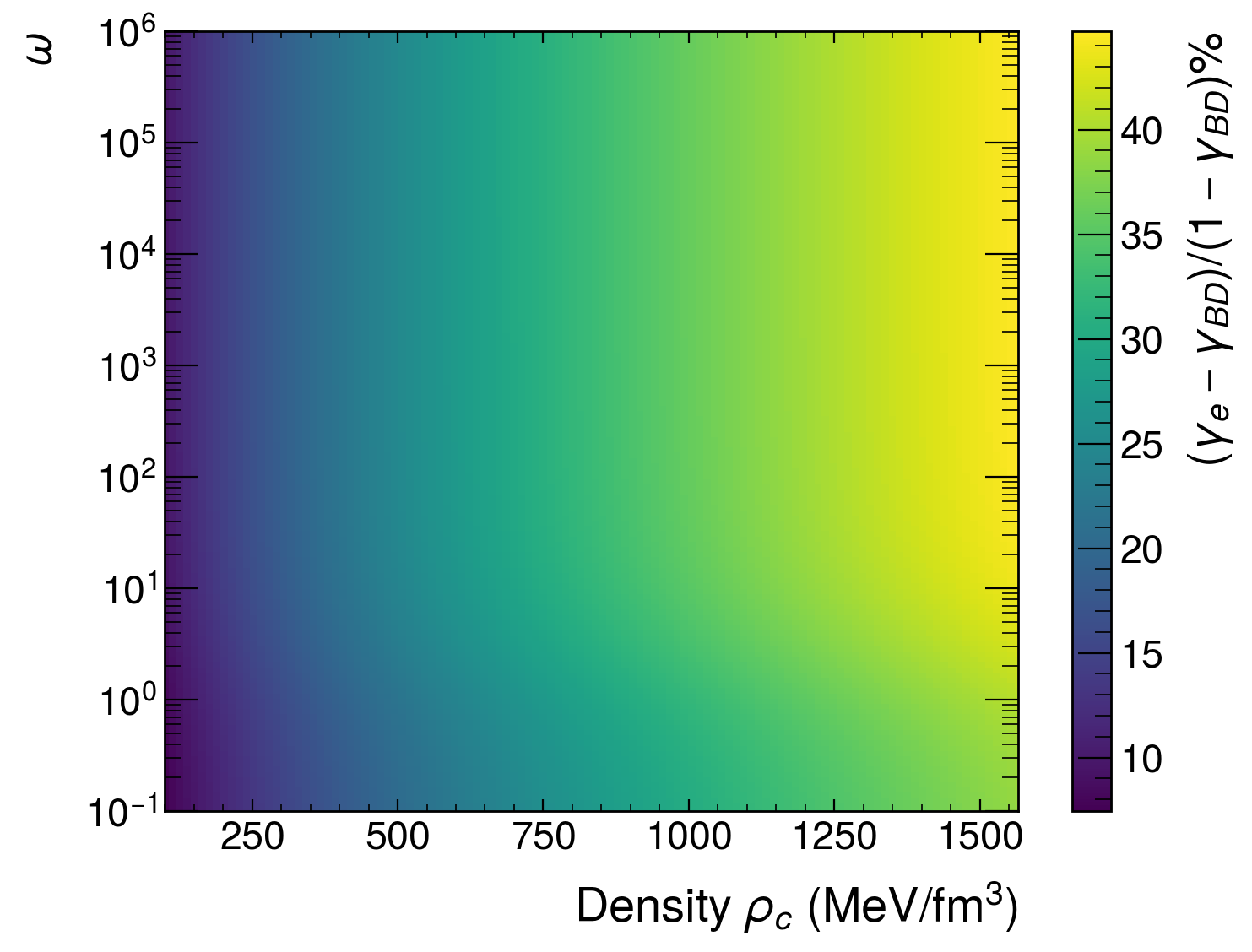}
\caption{Plot of the relative difference between the exact and standard post-Newtonian parameters $\gamma$ for 150 different values of $\omega$ and central densities. The values of $\omega$ ranges from $10^{-1}$ to $10^6$ meanwhile the values of central density ranges from $100$ MeV/fm$^3$ to 1570 MeV/fm$^3$.  This limit comes from the fact that we want the speed of sound in the neutron star to be less that the conservative limit considered of $c/\sqrt 3$  \citep{bedaque:2015pr}. In color shades is represented the relative deviation in percent of both $\gamma$ definitions.}
 \label{plot:relative}
	\end{center}
\end{figure}

\begin{figure}[h]
	\begin{center}
		\includegraphics[scale=0.45]{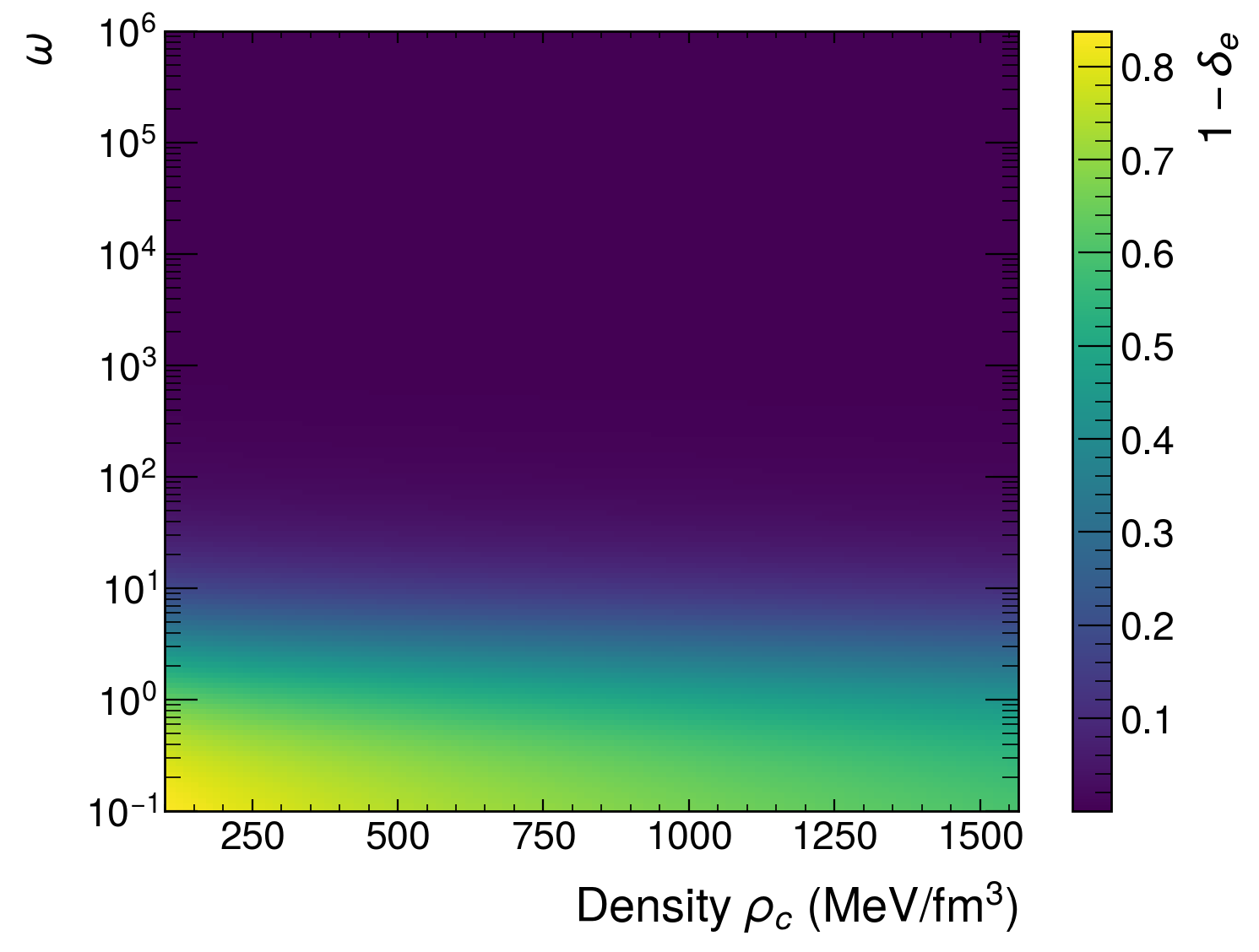}
\caption{Plot of $1-\deltae$ for 150 different values of $\omega$ and central densities. The values of $\omega$ ranges from $10^{-1}$ to $10^6$ meanwhile the values of central density ranges from $100$ MeV/fm$^3$ to 1570 MeV/fm$^3$. This limit comes from the fact that we want the speed of sound in the neutron star to be less that the conservative limit considered of $c/\sqrt 3$  \citep{bedaque:2015pr}. In color shades is represented $\deltae$.}
 \label{plot:deltae}
	\end{center}
\end{figure}

\begin{figure}[h]
	\begin{center}
		\includegraphics[scale=0.45]{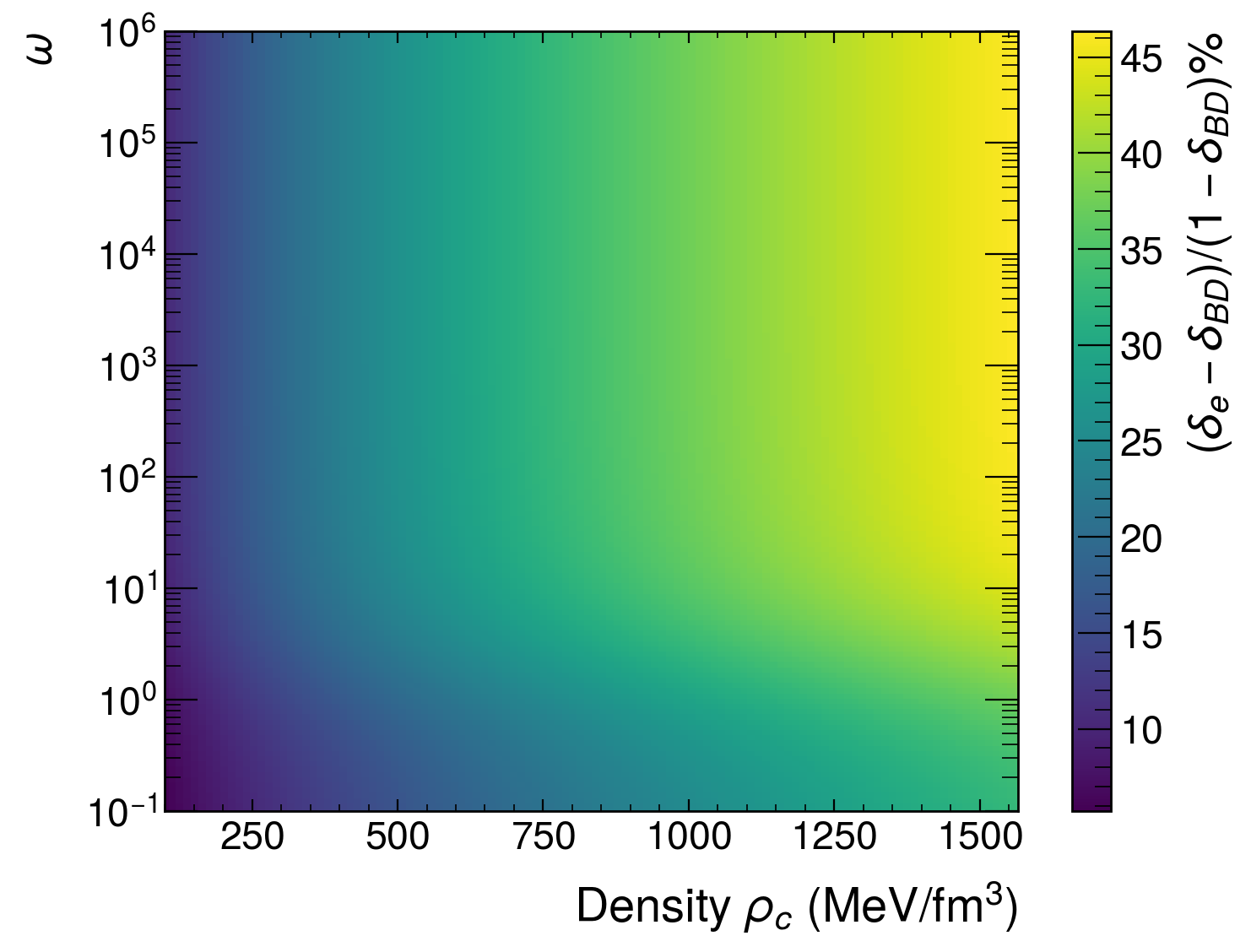}
\caption{Plot of the relative difference between the exact and standard post-Newtonian parameters $\delta$ for 150 different values of $\omega$ and central densities. The values of $\omega$ ranges from $10^{-1}$ to $10^6$ meanwhile the values of central density ranges from $100$ MeV/fm$^3$ to 1570 MeV/fm$^3$. This limit comes from the fact that we want the speed of sound in the neutron star to be less that the conservative limit considered of $c/\sqrt 3$  \citep{bedaque:2015pr}. In color shades is represented the relative deviation in percent of both $\delta$ definitions.}
 \label{plot:delta_relative}
	\end{center}
\end{figure}

In Fig. \ref{plot:edd_m} we see that the deviation from unity of $\gammae$ decreases with increasing $\omega$ and also with increasing energy density.
This is expected for increasing $\omega$, but it is somewhat counterintuitive for increasing central energy densities, given that $\Box \Phi \propto T$. But let us recall that the scalar field equations reads

\be \label{eq:unsourced}
\Box \phi = \frac{8\pi}{3+2\omega}\, ( - \rho + p_\parallel +2p_\perp ),
\ee

such that

\be \label{eq:unsourced_w}
\Box \phi \approx \frac{8 \pi \rho}{3+2\omega}\, \left( - 1 +  \frac{3p}{\rho} \right).
\ee

Therefore, because $P \rightarrow \rho/3$ for the most compact objects for which the speed of sound is close to $c/\sqrt{3}$,\footnote{Let us recall that the speed of sound $c_s$ is defined locally as $(c_s/c)^2=\partial P / \partial \rho$, such that $c_s = ~c/\sqrt{3}$ when $P=\rho/3$. Let us also recall that, following \cite{bedaque:2015pr}, we limited our integrations to cases that are such that $c_s = ~c/\sqrt{3}$.} the term in the parenthesis becomes closer to zero with an increasing central density.
This explains why, somewhat paradoxically, smaller deviations from General Relativity are found for the most compact objects. Overall, the deviation from unity is found to be up to 44.1\%.


In Fig. \ref{plot:relative}, one can see that, as the density goes up, more and more difference appears between the exact and standard post-Newtonian parameters $\gamma$. Indeed, for a given value of $\omega$, the value of $\gamma_{BD}$ stays the same, whereas the one of $\gammae$ approaches the General Relativity value---because of the cancellation in Eq. (\ref{eq:unsourced_w}) in the $P\rightarrow \rho/3$ limit. Hence, the relative variation increases with density. The relative variation is found to reach more than 44.7\%.

In Fig. \ref{plot:deltae}, we see the same behavior for $1-\deltae$, as the scalar field is less and less sourced with increasing $\omega$ and increasing $\rho$. The deviation from unity is found to be up to 83.7\%.

Lastly, in Fig. \ref{plot:delta_relative}, one can see that, as the density goes up, more and more difference appears between exact and standard post-Newtonian parameters $\delta$. For a given value of $\omega$, the value of $\delta_{BD}$ stay the same whereas the one of $\deltae$ approaches the General Relativity value---because of the cancellation in Eq. (\ref{eq:unsourced_w}) in the $P\rightarrow \rho/3$ limit. Hence the relative variation increases with density. The relative variation is found to reach 46.3\%.

The code used to generate all the figures presented in this manuscript is freely available on GitHub \cite{thomascode}. 


\subsection{Verification by comparing to an external numerical solution}\label{sec:verification}

In order to verify our results we will use an exact analytical solution derived in \cite{arruga:2021ep}. The method is to start from a general Janis-Newman-Winicour metric and conformally transform it to the Jordan frame. Thanks to what we can recover the Tolman-Oppenheimer-Volkoff metric. In what follows we explain how this works. \\

The Brans-Dicke action in the Einstein frame reads
\be
    S_{BD} = \frac{1}{2 \Tilde \kappa} \int d^4_{\t g}x \left[ \t R -2 \left( \partial \varphi \right)^2 + \t \L_m \left( \varphi, g_{\mu \nu}; \Psi\right) \right]
\ee
with the field definition\footnote{This normalization is different from the one used by \cite{chauvineau:2024prd}, and was chosen to be consistent with other previous works \cite{garfinkle:1991pr,holzhey:1992nb,minazzoli:2021ej,wavasseur:2025gg}.}
\be
d\varphi = \pm \frac{\sqrt{3 + 2\omega}}{2} d \ln{\phi}.
\ee

A class of vacuum spherical solutions is given by the Janis-Newman-Winicour metric:
\begin{equation}
\begin{aligned}
    ds^2 =& - \left( 1 - \frac{k}{r}\right)^d dt^2 \\
    &+\left( 1-\frac{k}{r}\right)^{-d} dr^2 + r^2 \left(1 - \frac{k}{r}\right)^{1-d} d\Omega^2,
\label{eq:JNW_metric}
\end{aligned}
\end{equation}
and the scalar field is given by
\be 
e^\varphi = \left(1 - \frac{k}{r} \right)^\frac{1 - d}{2\alpha},
\label{eq:JNW_scalar}
\ee
with $\alpha^2 = \frac{1 - d^2}{1+ d^2}$. Here, $d$ ranges between $[-1;0[ \cap ]0;1]$ and $d\Omega^2 = d\theta^2 + \sin^2(\theta) d\phi^2$ is the 2-sphere element. This writing of the Janis-Newman-Winicour solution can already be found in \cite{T_Damour_1992}.

For later convenience, it is useful to rewrite the solution using

\be \label{eq:alpha}
d = \frac{1-\alpha^2}{1+\alpha^2},
\ee

so that, one obtains

\begin{equation}
\begin{aligned}
    ds^2 = - \left( 1 - \frac{k}{r}\right)^{\frac{1-\alpha^2}{1+\alpha^2}} &dt^2 +\left( 1-\frac{k}{r}\right)^{-\frac{1-\alpha^2}{1+\alpha^2}} dr^2\\
    & + r^2 \left(1 - \frac{k}{r}\right)^{\frac{2 \alpha^2}{1+\alpha^2}} d\Omega^2.
\label{eq:JNW_metric}
\end{aligned}
\end{equation}

With $\alpha \in \mathbb{R}-\{1\}$, and 

\be 
e^\varphi = \left(1 - \frac{k}{r} \right)^\frac{\alpha}{1+\alpha^2}.
\label{eq:JNW_scalar}
\ee

We can now perform the inverse conformal transformation to set this Janis-Newman-Winicour metric into the Jordan frame to recover Tolman-Oppenheimer-Volkoff metric to study this solution numerically. The inverse transformation is given by 

\begin{equation}
    g_{\alpha \beta } =  e^{ \mp \xi \varphi}\t{g}_{\alpha \beta},
\label{eq:conformal_transf}
\end{equation}
with $\phi = e^{\mp \xi \varphi}$ in which $\xi$ is the conformal factor given by $\xi = \frac{2}{\sqrt{3 + 2\omega}}$ in the context of the Brans-Dicke theory. The resulting metric reads

\be
ds^2 = - adt^2 + b d\rho^2 + \rho^2 d\Omega^2.
\label{eq:tov_metric}
\ee

With

%

\begin{subequations}\label{eq:JNW-TOV}
\begin{eqnarray}
a &=& \left( 1 - \frac{k}{r}\right)^{\frac{1-\alpha^2 \mp \alpha \xi }{1+\alpha^2}}, \\
b &=& \left(\frac{dr}{d\rho}\right)^2 \left( 1-\frac{k}{r}\right)^{\frac{\alpha^2 - 1 \mp \alpha \xi}{1+\alpha^2}},  \\
\rho &=& r \left(1 - \frac{k}{r}\right)^{\frac{2\alpha^2 \mp \alpha \xi}{2\left(1+\alpha^2\right)}}, \\
\phi &=&  \left(1 - \frac{k}{r}\right)^{\pm \frac{\alpha \xi}{1+\alpha^2}}.
\end{eqnarray}
\end{subequations}

Hence, using conformal transformation of the Brans-Dicke theory, starting from the Janis-Newman-Winicour metric, we recovere Tolman-Oppenheimer-Volkoff metric. We have thus two ways to compute the exact parameters.\bigskip

\paragraph{\textbf{Computing exact parameters using numerical evaluation.}}

This is the method described in Sec. \ref{sec:exactPNBD}. In this case we explicitly use numerical evaluation of exact parameters\footnote{Let us note that while we use different radial coordinate systems in Sec. \ref{sec:BD} and Sec. \ref{sec:gammaBD}, the numerical evaluation of all the relevant quantities is invariant under this change, as emphasized in Sec. \ref{sec:invR}.} (equations (\ref{eq:exact_param_theta_BD})) by integrating numerically Eq. (\ref{eq:tointegrate}). \bigskip

\paragraph{\textbf{Computing exact parameters using external solution.}}

Another way comes from the evaluation of the parameter $d = \frac{1-\alpha^2}{1+\alpha^2}$ of the Janis-Newman-Winicour solution, from the external part of numerical solutions. Indeed using Tolman-Oppenheimer-Volkoff solution given by Eq. (\ref{eq:tov_metric}), one obtain equation that governs the scalar field, in the vacuum limit 
\be
\frac{\Ddot{\phi}}{\dot{\phi}} = \frac{1}{2} \left[\frac{\dot{b} }{b} - \frac{\dot{a}}{a} - \frac{4}{\rho} \right].
\ee
Then, by integration, one obtains

\be
\dot{\phi} = \frac{C}{\rho^2} \sqrt{\frac{b}{a}},
\ee
with $C$ a constant of integration. Finally using Eq. (\ref{eq:JNW_scalar}) and Eq. (\ref{eq:JNW-TOV}), by computing the limit when $\rho \rightarrow \infty$, one obtains

\be
\alpha = \frac{-\frac{2}{\sqrt{3}} \left( 1+C\right) - \sqrt{ \Delta}}{2},
\ee
with $\Delta$ given by $\Delta = \left( \frac{2}{\sqrt{3}} \left( 1+C \right) \right)^2 +4$.

Thanks to which we can evaluate exact post-Newtonian parameters, in Brans-Dicke theory (i.e., by switching from $\alpha$ to $d$ with Eq. (\ref{eq:alpha}) and applying $\omega = cst$ to equations (\ref{eq:exact_param})) given by

\begin{subequations}\label{eq:exact_param_BD_alpha}
\begin{align}
\beta_{\text{exact}} 
&= 1,\\
\gamma_{\text{exact}} \label{eq:exact_param_BD_gammae}
&= \frac{ d - \sqrt{ \frac{1 - d^2}{3+2\omega_0} } }{ d + \sqrt{ \frac{1 - d^2}{3+2\omega_0} } }, \\
\delta_{\text{exact}}
&= \begin{aligned}[t]
   \frac{4}{3} & \gamma_{\text{exact}}^2 -\\
   & \frac{1}{3} \left( d + \sqrt{ \frac{1 - d^2}{3+2\omega_0} } \right)^{-2}.
   \end{aligned}
\end{align}
\end{subequations}

The method that deduces the Janis-Newman-Winicour parameter $\alpha$ from the external part of numerical Tolman-Oppenheimer-Volkoff solutions was originally presented in  \citep{arruga:2021ep}.\bigskip

\paragraph{\textbf{Cross-checking results}.}

Hence we have two methods to compute the exact parameters, such that we can cross-check their results. One that fits the Janis-Newman-Winicour parameter $\alpha$ (or $d$) to the external part of numerical solutions, and the other that numerically computes analytical expressions that depends on $a$, $b$, $\rho$ and $P$ originally presented in \cite{chauvineau:2024prd}---see Sec. \ref{sec:exactPNBD}.

We have verified that the highest relative deviation in absolute value, computed by our code between both $\gammae$ estimations are below 0.07\% and at most of $0.3\%$ for $\deltae$.\\

This notably is an independent test of the validity of the analytical results presented in \cite{chauvineau:2024prd}.

\subsection{Discussion on Bergmann-Wagoner-Nordvert theories}\label{sec:discGenST}

General Bergmann-Wagoner-Nordvert theories are known to have non-perturbative strong-field effects that enhance the scalar charge of compact objects \cite{damour:1993pl}. This process is known as the `\textit{spontaneous scalarization}' of compact objects \cite{damour:1996pr}. It has notably been used in order to constrain scalar-tensor theories, notably with white dwarf-neutron star systems, through the fact that a strong dipolar emission of gravitational waves is expected from scalarized compact objects, whereas no sign of dipolar emission is found in the observational data to date \cite{freire:2012mn}.\\

Therefore, since the exact post-Newtonian parameters are non-perturbative quantities, one can expect spontaneous scalarizations to induce a significant deviation of the exact post-Newtonian parameters from the standard post-Newtonian parameters. This should then be compared to the body-dependent parameters $\alpha_A$ and $\beta_A$ in the Damour-Esposito-Farèse parametrization of Scalar-Tensor theories of gravity \cite{damour:1992cq,damour:1996pr}, which are also sensitive to non-perturbative effects---unlike standard post-Newtonian parameters. This is the topic of the following sub-section.\\

Let us stress, however, that spontaneous scalarization cannot occur in either of the two models considered in the present paper, since in both cases $\omega(\phi)$ is constant---even zero, in the case of Entangled Relativity \cite{ludwig:2015pl}.

\subsection{Linking Damour-Esposito-Farèse non-perturbative parameter to $\gammae$}
\label{sec:DEFvsGammae}

One can expect a direct link between the scalar charge of compact object and exact post-Newtonian parameters. Non-perturbative strong field effects have notably been studied in binary system by \cite{Damour_1996}. Starting from a scalar-tensor theory in Jordan frame, an arbitrary coupling function $A(\varphi)$ arise when changing to the Einstein frame as $\t g_{\mu \nu} = A^2(\varphi) g_{\mu \nu}$. Following \cite{Damour_1996}, in the far scalar field region we have

\be
\varphi = \varphi_0 - \frac{m_A \alpha_{DEF}}{r} + \mathcal O(r^{-2}),
\ee
in which $\varphi$ is the scalar field and $\varphi_0$ its value at infinity. $m_X$ represent the mass of the body $X$ and $\alpha_{DEF}(\varphi) = \frac{\partial \ln{A(\varphi)}}{\partial \varphi}$. The subscript \textit{DEF} stands for \textit{Damour and Esposito-Farèse}. This should be compared to the expansion of the scalar field equation (\ref{eq:JNW_scalar}) of the Janis-Newman-winicour solution, in the far field region. This one reads

\be
\varphi = \varphi_0 - \frac{\alpha_{JNW}}{1-\alpha_{JNW}^2} \frac{2m_A}{r}+ \mathcal O(r^{-2}).
\ee
The subscript \textit{JNW} stands for \textit{Janis-Newman-Winicour}. By identification, we have

\be
\alpha_{DEF} = \frac{2 \alpha_{JNW}}{1 - \alpha_{JNW}^2}.
\ee
Hence we can derive a link between the scalar charge of Damour-Esposito-Farèse and the one of Janis-Newman-Winicour. Using Eq. (\ref{eq:gammae}) and keeping in mind Eq. (\ref{eq:alpha}), one finally obtains a formulation depending on the exact Post-Newtonian parameter, for the Brans-Dicke theory,

\be\label{eq:alphaDEF-alphaJNW}
\alpha_{DEF} =  \sqrt{3+2\omega_0}\left( \frac{1 - \gammae}{1+\gammae}\right),
\ee
or, alternatively,

\be
\gammae = \frac{\sqrt{3+2\omega_0} - \alpha_{DEF} }{\sqrt{3+2\omega_0} + \alpha_{DEF} }.
\ee

Let us stress that the relation between $\gammae$ and $\alpha_{DEF}$ follows directly from their respective definitions in the far-field region and is therefore fully general.

\section{Exact parameters in Entangled Relativity}
\label{sec:ER}

The aim of this section is to provide a brief overview of the theory of Entangled Relativity and then to examine the standard post-Newtonian expansion of the theory and derive an exterior solution of it, which will be useful in the next section for the numerical evaluation of $\gammae$ and $\deltae$ within the solar system and neutron stars.

\subsection{The theory}
\label{sect:ER_th}

One of the motivations of this theory comes from the fact that by a simple modification of the Einstein-Hilbert action one can immediately satisfy Mach's principle\footnote{Given that Mach never formulated a \textit{Mach's Principle}, there exist many different versions of what Mach's principle could mean \cite{book_mach_principle}. Here we refer to the principle that Einstein used to construct his theory of General Relativity and which he named \textit{Mach's principle} in \cite{einstein:1918an}, see also \cite{hoefer:1995cf,minazzoli:2024pn}.} while being more economical in terms of fundamental constants than General Relativity and still recovering the phenomenology of General Relativity in some limit. Mach's principle---or principle of relativity of inertia---states that movement, and thus inertia, cannot be defined with respect to nothing. In relativistic theories, inertia is defined through the metric tensor. But General Relativity admits vacuum solutions, thereby automatically violating Mach's principle. 

In Entangled Relativity, instead of the standard additional coupling of General Relativity between matter and spacetime in the action, they are coupled non-linearly---see Eq. (\ref{eq:ERS}). Because of that, the theory cannot even be defined without matter fields. Moreover, the field equations are such that vacuum solutions are ill-defined, and therefore prohibited---although it has been shown that all the vacuum solutions of General Relativity are limits of solutions of Entangled Relativity when the density of matter fields tends toward zero \cite{minazzoli:2025ej,minazzoli:2025pl}. 

To see this more clearly, let us define the theory from its action \citep{ludwig:2015pl,minazzoli:2018pr,minazzoli:2022ar,chehab:2025cq}

\be
    S_{ER} \propto \int d^4x \sqrt{-g} \frac{\Lm^2}{R},\label{eq:ERS}
\ee
where $R$ is the usual Ricci scalar that is constructed upon the metric tensor $g$, $ \mathrm{d}^4_g x := \sqrt{-|g|}  \mathrm{d}^4 x$ is the spacetime volume element, with $|g|$ the metric $g$ determinant, and $\L_m$ is the Lagrangian density of matter. As one can see in Eq. (\ref{eq:ERS}), if $\L_m(f,g) = \emptyset$ in the action, the theory cannot be defined at all.

It has been shown in \cite{Minazzoli_2018, ludwig:2015pl,minazzoli:2022ar} that there is a one to one correspondence at the classical level between the action of Entangled Relativity and an Einstein-dilaton theory (provided that $\Lm \neq \emptyset$)  with the following action

\be 
S_{Ed} \propto\int d^4x \sqrt{-g} \left( \frac{\phi R}{2 \t \kappa} + \sqrt{\phi} \Lm \right)\label{eq:ERS_eq}.
\ee
Here, $\tilde{\kappa}$ is a positive effective coupling constant between matter and geometry, with the same dimension as $\kappa$ from General Relativity and $\phi$ is a dimensionless scalar field. The classical equivalence between the original \textit{$f(R,\L_m)$} theory Eq. (\ref{eq:ERS}) and the \textit{Einstein-dilaton} theory Eq. (\ref{eq:ERS_eq}), comes from the fact that any non-linear algebraic function of the Ricci scalar in the action are equivalent to having an additional scalar degree-of-freedom with gravitational strength \citep{capozziello:2015sc,teyssandier:1983jm,jakubiec:1988pr}. To obtain the field equations, we apply the least action principle, i.e. : $\delta S_{Ed} = 0$. The metric field equation reads

\begin{equation}
\begin{aligned}
    R_{\mu \nu} -& \frac{1}{\phi} \partial_\mu \partial_\nu \phi + \frac{1}{\phi} \Gamma_{\mu \nu}^\lambda \partial_\lambda \phi \\
    & = \frac{\tilde{\kappa}}{\sqrt{\phi}} \left( T_{\mu \nu} - \frac{1}{2} T g_{\mu \nu} \right) + \frac{1}{2} \frac{\Box \phi}{\phi} g_{\mu \nu}, \label{eq:FE_ER}
\end{aligned}
\end{equation}

\be
\Box \phi = \frac{\t{\kappa} \sqrt{\phi}}{3} (T- \L_m). 
 \label{eq:FE_phi}
\ee
With $T$ the trace of the stress energy tensor. 

Interestingly, even if the action of the theory appears quite different from that of General Relativity, it recovers its phenomenology in many common situations. This comes from what has been called the \textit{intrinsic decoupling} \citep{minazzoli:2013pr}. This kind of decoupling takes place when one has $\Lm = T$. 

Indeed in Eq. (\ref{eq:FE_phi}) if $\Lm = T$, the scalar field is not sourced anymore and the field equation given by Eq. (\ref{eq:FE_ER}) can reduce to those of General Relativity without a cosmological constant. This $\Lm = T$ equality is, for example, a good approximation for a universe made of dust and electromagnetic radiation, just like our universe. 
\subsection{On-shell value of $\Lm$}

As explained in \cite{chehab:2025cq}, in some simple cases, the on-shell matter Lagrangian can be obtained directly from the field equations. For instance, for a charged black hole, one finds in natural units that the on-shell matter Lagrangian is $\Lm = E^2/2$, where $E$ is the norm of the electric field \citep{minazzoli:2021ej,wavasseur:2025gg}.\\

For more general systems, however, and especially for realistic astrophysical bodies, such a first-principles derivation is usually out of reach. One is therefore led to adopt assumptions whose validity may depend on the physical system under consideration. As a consequence, although Entangled Relativity has no free parameter in its definition, its phenomenology may still depend on the assumption made for the on-shell value of the matter Lagrangian, unlike General Relativity.\\

In this work, we shall assume that matter can be described as a perfect fluid and that the rest-mass energy density $\rho_r$ is conserved, namely
\be
\nabla_\sigma (\rho_r u^\sigma)=0.
\ee
Under this assumption, the on-shell Lagrangian is $\Lm=-\rho$, where $\rho$ denotes the total energy density, defined by \citep{minazzoli:2012pr,arruga:2021pr}
\be \label{eq:deftotenergy}
\rho = \rho_r\left(c^{2}-\frac{P}{\rho_r}+\int \frac{\mathrm{d} P}{\rho_r}\right).
\ee
In that case, the scalar field has been referred to as a \textit{pressuron} \citep{minazzoli:2014pr}, since it is sourced only by pressure---as it can be seen from Eq. (\ref{eq:phiP}).\\

Another possibility that has been advocated in the literature is to identify the on-shell matter Lagrangian with the trace of the stress-energy tensor, namely $\Lm=T$ \citep{avelino:2018pr,avelino:2022pr,pinto:2025ar}. This is a distinct assumption from the one adopted here. If one further assumes that the usual definition of the total energy density, Eq. (\ref{eq:deftotenergy}), still applies, then the rest-mass energy density is in general no longer conserved, since one has \citep{arruga:2021pr}
\be \label{eq:nonconsrtho_P}
\nabla_\sigma (\rho_r u^\sigma)=-\frac{3}{2} P \frac{\rho_r}{\rho+P} u^{\sigma} \partial_{\sigma} \ln\phi
\qquad \textrm{if } \Lm=T.
\ee
That being said, if $\Lm=T$ for the compact bodies considered below, then one can still recover General Relativity exactly because the right-hand side of Eq. (\ref{eq:FE_phi}) vanishes identically and $\phi$ can be constant. In that situation, Eq. (\ref{eq:nonconsrtho_P}) also implies that the rest-mass energy density remains conserved, simply because one has $\partial_\sigma \phi =0$ in that case. Let us stress that the fact that $\nabla_\sigma (\rho_r u^\sigma)=0$ can hold both for $\Lm=-\rho$ and for $\Lm=T$ is a distinctive feature of theories exhibiting \textit{intrinsic decoupling} \citep{minazzoli:2013pr,minazzoli:2026pl}, within the broader class of theories with non-minimal scalar-matter couplings \citep{harko:2013pr}.\\

Since the choice $\Lm=T$ leads exactly back to General Relativity in the Solar System, for white dwarfs, and for neutron stars, we shall restrict the following analysis to the case $\Lm=-\rho$.\\

Let us finally emphasize that determining which on-shell matter Lagrangian is the most appropriate for a given astrophysical body---or whether several different on-shell Lagrangians should coexist depending on the fields involved and their physical state---remains an open question.
\subsection{Usual post-Newtonian derivation}
\label{sec:pN}

The usual post-Newtonian derivation is a perturbative scheme that assumes that the motion of Solar System objects is such that $v/c \sim \sqrt{Gm/(rc^2)} \ll 1$ and that $T^{\alpha \beta} = O(c^2,c^1,c^0)$ \cite{damour:1991pr}. It is in particular perfectly relevant in the weak-field limit of the solar-system, but becomes less relevant in the strong-field regime of compact objects such as neutron stars and black-holes. Assuming $\L_m = -\rho$ \cite{minazzoli:2012pr,arruga:2021pr}, Eq. (\ref{eq:FE_phi}) becomes
\be\label{eq:phiP}
\Box \phi =  \t\kappa \sqrt{\phi} P, 
\ee
where $P$ is the pressure of the fluid sourcing the gravitational field equations.
In either the harmonic or standard post-Newtonian gauges, the first order post-Newtonian metric can be parametrized by a potential $w$, a vector potential $w^i$ and the post-Newtonian parameters $\gamma_{PN}$ and $\beta_{PN}$ \citep{minazzoli:2013pr}
\begin{subequations}\label{eq:PPNmetric}
\begin{eqnarray}
	g_{00}&=&-1+2\frac{w}{c^2}-2\beta_{PN}\frac{w^2}{c^4}+\mathcal O(1/c^6),\\
	g_{0i}&=&-2(1+\gamma_{PN})\frac{w^i}{c^3}+\mathcal O(1/c^5),\\
	g_{ij}&=&\delta_{ij}\left(1+2\gamma_{PN}\frac{w}{c^2}\right)+\mathcal O(1/c^4). \label{eq:ssmetric}
\end{eqnarray}
\end{subequations}
Injecting this metric in the field equations~(\ref{eq:FE_ER}) and~(\ref{eq:phiP}) results in 
\be
\gamma_{PN}=\beta_{PN}=1,
\ee 
and
\begin{subequations}
\begin{eqnarray}
w&=&w_{GR} - \frac{1}{c^2} G \int  \frac{P(\bold{x}')d^3 x'}{|\bold{x}-\bold{x}'|} +\mathcal O(1/c^4),\nonumber \\
&=:& w_{GR} + \frac{1}{c^2} \delta w+\mathcal O(1/c^4),\label{eq:pot}\\
w^i&=&w^i_{GR} +\mathcal O(1/c^2),\label{eq:poti}
\end{eqnarray}
\end{subequations}
where $w_{GR}$ and $w^i_{GR}$ are the expressions of the potentials predicted by General Relativity. From Eq.~(\ref{eq:PPNmetric}) and Eq.~(\ref{eq:pot}), one can see that the metric characterizing the considered theory differs from the General Relativity metric at the first post-Newtonian level by a $1/c^4$ term in the temporal component of the metric. The solution of the scalar-field equation~(\ref{eq:phiP}) can be written as
\begin{equation}\label{eq:phiSol}
\frac{\delta \phi}{\phi_0}= - 2 \delta w+\mathcal O(1/c^2)  ,
\end{equation}
where $\delta \phi:= c^4 (\phi- \phi_0)$. In other words, at leading order, one has
\begin{subequations}\label{eq:PNAB}
\bea
\label{eq:A_PN}
    A &&= 1 - \frac{2\omega}{c^2} + O(c^{-4}),\\
    B &&= 1 + \frac{2\omega}{c^2} + O(c^{-4}),
\label{eq:B_PN}
\eea
\end{subequations}
where
\be
\omega = G \int  \frac{\rho(\bold{x}')d^3 x'}{|\bold{x}-\bold{x}'|} + O(c^{-4}).
\ee

\subsection{Deriving an exterior solution of Entangled Relativity stars}

The goal of this part is to derive an exterior solution for Entangled Relativity stars using the same methodology as in the Brans-Dicke case, briefly explained in Sec \ref{sec:BD}, used by \cite{chauvineau:2024prd}. The logic is exactly the same, only the field equations are slightly changed between the two cases.\\

Hence we start in Einstein frame with the Janis-Newman-Winicour metric \citep{janis:1968prl} that we re-expressed in the Jordan frame thanks to a conformal transformation. One considers the following rescaling of the scalar field

\be \label{eq:CT_ER}
d\varphi = \frac{\sqrt{3}}{2} d\ln{\Phi},
\ee
while the metric solution reads

\be
\label{eq:JNW_JF_ER}
ds^{2} = -A dt^{2} + B dr^{2} + C d\Omega^{2},
\ee
where the function A, B and C are given by 
\be\label{eq:JNW_ER}
\begin{cases}
A = -\frac{1}{\Phi}\left( 1 - \frac{k}{r} \right)^{d}, \\[1.2em]
B = \frac{1}{\Phi}\left( 1 - \frac{k}{r} \right)^{-d}, \\[1.2em]
C = \frac{r^2}{\phi} \left( 1 - \frac{k}{r} \right)^{1-d}, \\[1.2em]
\varphi = -\frac{\sqrt{1 - d^2}}{2} \ln\left(1 - \frac{k}{r}\right), \\[1.2em]
\end{cases}
\ee
in which $k$ is linked to the mass by $k = \frac{2m}{d}$ and $d$ is the scalar charge. 
The stationarity and spherical symmetry of the solution imposes an energy-momentum tensor of the form 
\be
T^\mu_\nu = \text{diag}(-\rho, P_\parallel, P_\perp, P_\perp),
\label{eq:stress_energy}
\ee
where $\rho$ is the energy density, $ P_\parallel$ and $P_\perp$ are respectively the parallel and perpendicular pressure. These are dependent functions of $r$. 

To derive the exterior solution, only the (00) component of the field equation Eq. (\ref{eq:FE_ER}) and the scalar field equation Eq. (\ref{eq:FE_phi}) are relevant. They read

\begin{equation} \label{eq:FE00ER}
\begin{aligned}
   \left( \frac{\phi C A_{,r}}{\sqrt{AB}} \right)_{,r}  =& \frac{\tilde{\kappa} \sqrt{\phi}}{3} \left( -4 T^0_0 + 2 T^i_i + \Lm \right) r^2 \sqrt{AB^3},
\end{aligned}
\end{equation}

\begin{equation}\label{eq:FEphiER}
\begin{aligned}
    \left( \sqrt{\frac{A}{B}} C \phi_{,r} \right)_{,r} = & \frac{\tilde{\kappa} \sqrt{\phi}}{3} \left( T^0_0 + T^i_i - \Lm \right) r^2 \sqrt{AB^3},
\end{aligned}
\end{equation}
where $X_{,r} := \frac{\partial X}{\partial r}$. As the left-hand side of these equations is in exact derivative form, it can easily be integrated from the star's center to a distance $r$ superior to the star's radius $r_*$. Before doing so, let us define some integrals

\begin{subequations}
\bea
\label{eq:defeps}
     E^* &&:= \int_0^{r_{*}} dr\frac{\tilde{\kappa} \sqrt{\phi}}{3} \left( -T^0_0 \right) r^2 \sqrt{AB^3},\\
\label{eq:defp}
     P^* &&:= \int_0^{r_{*}}dr \frac{\tilde{\kappa} \sqrt{\phi}}{3} T^i_i r^2 \sqrt{AB^3},\\
     \mathcal{L}_m^* &&:= \int_0^{r_{*}}dr \frac{\tilde{\kappa} \sqrt{\phi}}{3} \mathcal{L}_m r^2 \sqrt{AB^3}.
\label{eq:deflm}
\eea
\end{subequations}
Hence Eq. (\ref{eq:FE00ER},\ref{eq:FEphiER}) becomes

\be
    \frac{\phi C A_{,r}}{\sqrt{AB}} = 4E^* +2P^* + \Lm^*,\\
\ee

\be
    \sqrt{\frac{A}{B}} C \phi_{,r} = -E^* +P^* - \Lm^*.
\ee
Now, by solving with the metric Eq. (\ref{eq:JNW_ER}), one finds

\begin{subequations}
\bea
    dk + k\frac{\sqrt{1-d^2}}{\sqrt{3}} &&= 4E^* + 2P^* + \Lm^*,\\
    k \frac{\sqrt{1-d^2}}{\sqrt{3}}  &&= E^* - P^* +\Lm^* ,
\eea
\end{subequations}
and then one finds

\begin{subequations}
\bea
    k &&= 3 \sqrt{ \left( E^* +P^* \right)^2 + \frac{\left( E^* -P^* + \Lm^* \right)^2}{3} },\\
    d  &&= \frac{3\left(E^* + P^*\right)}{k}.
\eea
\end{subequations}
As in the Brans-Dicke case, one can formulate a parameter $\Theta = \frac{P^* }{E^*}$ and hence rewrite $k$ and $d$ as

\begin{subequations}\label{eq:d_k_ER}
\bea
    k &&= 3 E^* \sqrt{ \left( 1+ \Theta \right)^2 + \frac{\left(1 - \Theta + \frac{\Lm^*}{E^*} \right)^2}{3}},\\
    d  &&= \frac{1+\Theta}{ \sqrt{ \left( 1+ \Theta \right)^2 + \frac{\left(1 - \Theta + \frac{\Lm^*}{E^*} \right)^2}{3}}}.
\eea
\end{subequations}
Now one can inject Eq. (\ref{eq:d_k_ER}) into Eq. (\ref{eq:exact_param}) as the latter has been obtained through metric expansion in \cite{chauvineau:2024prd}.\footnote{Let us stress once again that the parameters and their analytical expressions are invariant under a change of the radial coordinate---see Sec. \ref{sec:invR}.} In the case of Entangled Relativity, $\omega = 0$. Finally one obtains a formula for the exact parameters in Entangled Relativity.

\begin{subequations}\label{eq:gamma_ER}
\begin{align}
\betae  &= 1,\\
 \gammae &= \frac{1 + 2 \Theta - \frac{\mathcal{L}_m^*}{2 E^*} }{2 + \Theta + \frac{\mathcal{L}_m^*}{2E^*}},\\
\deltae&
\begin{aligned}[t]
    &=\frac{4}{3}  \gammae^2 -\\
    &\frac{1}{3} \left( \frac{4+2 \Theta + \frac{\Lm^*}{E^*}}{\sqrt{9\left(1+\Theta \right)^2 + 3 \left( 1-\Theta + \frac{\Lm^*}{E^*} \right)^2} }\right)^{-2} .
   \end{aligned}
\end{align}
\end{subequations}
One can especially see that the type of matter will play a crucial role in the resulting curvature because $\Lm^*$ appears in the computation of the exact post-Newtonian parameters. For instance, for an electromagnetic radiation one has $\Lm = T = 0$, whereas for a magnetic field, one has $T\neq \Lm \propto B^2$, where $B$ is the magnetic field.
If one considers the example of a dust fluid (i.e. $P=0$ and $\Lm = - \rho$), one recovers General Relativity with $\gamma_{exact} = 1$.
This is an expected result, since one has $\Lm = -\rho = T$ for a dust fluid, which implies that the scalar field is not sourced in Eq. (\ref{eq:FE_phi}).
\section{Numerical evaluation of exact parameters in Entangled Relativity}
\label{sec:gammaER}

In this section we will compute the numerical evaluations of ($\gammae,\betae, \deltae$) in the solar system and for neutron stars in Entangled Relativity.

\subsection{In the Solar System}

Although the exact post-Newtonian parameters have been derived in the strong-field limit, they can still be derived in the perturbative weak-field case of the usual post-Newtonian developments. In order to derive exact parameters, one needs to know the internal value of $A$, $B$ and $\phi$, which are given at leading order by Eq. (\ref{eq:PNAB}) and $\phi=\phi_0+O(c^{-4})$. Eq. (\ref{eq:defeps}-\ref{eq:defp}) therefore reduce at leading order to

\begin{equation}
\begin{aligned}
\label{eq:E*}
     E^* := \frac{\t \kappa}{3}& \int_0^{r_{*}} \rho r^2 \sqrt{\phi_0} dr\\ 
     &+ \frac{\t \kappa}{3} \int_0^{r_{*}} \rho r^2 \sqrt{\phi_0} \frac{2\omega_{int}}{c^2} dr+ O(c^{-4}),\\
\end{aligned}
\end{equation}

\begin{equation}
\begin{aligned}
\label{eq:p*}
     P^* := \t \kappa &\int_0^{r_{*}} Pr^2 \sqrt{\phi_0} dr\\ 
     &+ \t \kappa \int_0^{r_{*}} Pr^2 \sqrt{\phi_0} \frac{2\omega_{int}}{c^2} dr+ O(c^{-4}),
\end{aligned}
\end{equation}


where, assuming a spherical symmetry, one has
\begin{equation}
\label{eq:omega}
    \omega_{int}(r) =  \frac{4 \pi G}{r} \int_0^{r} \rho(r') r'^2 dr',
\end{equation}
where $\omega_{int}$ is the interior gravitational potential. To evaluate the numerical integration, one can use the model S of the sun ($\odot$) \citep{model_S} and the PREM model for the earth ($\oplus$) \citep{PREM}, as both give the pressure and density values of the considered bodies at a considered radius.  By computing  $ \Theta = \frac{P^*}{E^*}$ and using equations (\ref{eq:gamma_ER}), one finds, for $\Lm=-\rho$,
\begin{subequations}
\bea
\label{eq:gamma_sun}
    \gammae^{\odot}-1 && \approx 7.10^{-6},\\
    \gammae^{\oplus}-1 &&\approx 9.10^{-10},\\\nonumber\\
    \deltae^{\odot}-1 && \approx 2.10^{-5},\\
    \deltae^{\oplus}-1 &&\approx 2.10^{-9}.
\label{eq:gamma_earth}
\eea
\end{subequations}
The smallness of the values of the exact parameters for Solar System bodies  in Entangled Relativity comes from the \textit{intrinsic decoupling} mentioned in Sec. \ref{sect:ER_th}.
The codes that we used to compute these values are freely accessible on github at \cite{thomascode}. These predicted values are beyond our current measurement capabilities, but do not seem to be too far out of reach in the case of the Sun. Indeed, current constraints are $\gamma_{\odot} = 1 + (2.1 \pm 2.3)\times10^{-5}$ \citep{Bertotti2003}. However, at the $10^{-6}$ level of accuracy for the constraint on $\gamma$, because $\sqrt{Gm_\odot/(R_\odot c^2)} \sim 10^{-6}$, one can no longer use the leading order development of the propagation of electromagnetic waves and one has to use a full $c^{-4}$ development instead, in order to be fully consistent at the numerical level. This is beyond current best practices \cite{fienga:2024lr}, although several studies started to address this difficult question already \cite{ashby:2010cq,minazzoli:2011cq,deng:2012pr,linet:2013cq,hees:2014pr,linet:2016pr,cappuccio:2021cq,zschocke:2022pr}. 
\\

Of course, this is consistent with the perturbative approach of the usual post-Newtonian parameters, as what $\sqrt{Gm_\odot/(R_\odot c^2)} \sim 10^{-6} \sim \gammae^{\odot}-1$ means is that $\gammae-1 = O(c^{-2})$, and therefore can be neglected at the first post-Newtonian order. This fact changes in the case of ultra-relativistic strong-field compact objects, where $\gammae - 1$ can become somewhat significantly different from zero, as we will now see.

\subsection{For neutron stars}

By solving the modified Tolman-Oppenheimer-Volkoff equation, one can simulate compact objects in Entangled Relativity. Because it has been shown that Entangled Relativity may predict deviations near compact objects with respect to General Relativity \citep{chehab:2025cq, arruga:2021pr}, we decided to model neutron stars. We used a simple polytropic equation of state (as the goal is to capture the broad behavior of $\gammae$ and $\deltae$, exactly as explained for the Brans-Dicke case) given by 

\be \label{eq:eos}
P=K \rho^{\Gamma},
\ee
with $P$ the pressure, $\Gamma$ the adiabatic index given by $\Gamma = 5/3$ and a constant $K = 1.475 \times 10^{-3}$ (fm$^3$/MeV)$^{2/3}$ \citep{arbanil:2013pr,arruga:2021pr}. We conducted the simulation between 100 and 2000 MeV/fm$^3$ but we only kept the simulations that are such that the speed of sound inside the neutron star is less than the conservative limit considered of $c/\sqrt{3}$ \citep{bedaque:2015pr}. This one corresponds to a central density of 1578 MeV/fm$^3$ for the considered equation of state. 

The deviation from unity of $\gammae$ and $\deltae$ can be computed with equations (\ref{eq:gamma_ER}), and verified as for the Brans-Dicke case with equations (\ref{eq:exact_param}), with $\omega = 0$. Here again, both parameter have been calculated with the two methods described in sec. \ref{sec:verification}, and give the same results with a maximal relative difference of 0.01\% for $\gammae$ and of 0.02\% for $\deltae$. The resulting plot showing the behavior of $1 - \gammae$ and $1- \deltae$ are presented in Fig. \ref{plot:gamma_alpha} and Fig. \ref{plot:delta_ER} in the context of Entangled Relativity, where $\gammae$ and $\deltae$ are computed with equations (\ref{eq:gamma_ER}). 

As we can see in Fig. \ref{plot:gamma_alpha}, the deviation of $\gammae$ increases with the density.  The deviation ranges, in absolute value, from 4.7\% to 16.1\%.
Indeed, the denser the body, the larger the deviation, as it can be expected with Eq. (\ref{eq:phiP}). The denser the object, the more strongly the scalar field is sourced and hence more deviation from General Relativity should occur. As expected, regarding the behavior of $\deltae$, we can see on Fig. \ref{plot:delta_ER} that it follows the same trajectory as for $\gammae$. The deviation ranges, in absolute value, from 11.2\% to 40.1\%.

These results have been obtained without any free parameter but use one assumption : $\Lm = -\rho$. If $\Lm = T$ instead, then no deviation occurs relative to General Relativity, since in Eq. (\ref{eq:FE_phi}), the scalar field is not sourced and the field equation Eq. (\ref{eq:FE_ER}) becomes the one of General Relativity.

The equation and code to model neutron stars in Entangled Relativity has already been made and computed in \cite{arruga:2021pr}. The codes that generates all the figures presented in this manuscript, are freely available on github \cite{thomascode}. Given that the code integrates the equations of a somewhat recent theory, we use a first integral of the equations in order to further check the integration in Sec. \ref{app:validate}. We find that the constant from the first integral remains constant at the $10^{-11}$ level, thereby validating the integration at a very good level of accuracy.

\begin{figure}[h]
	\begin{center}
		\includegraphics[scale=0.45]{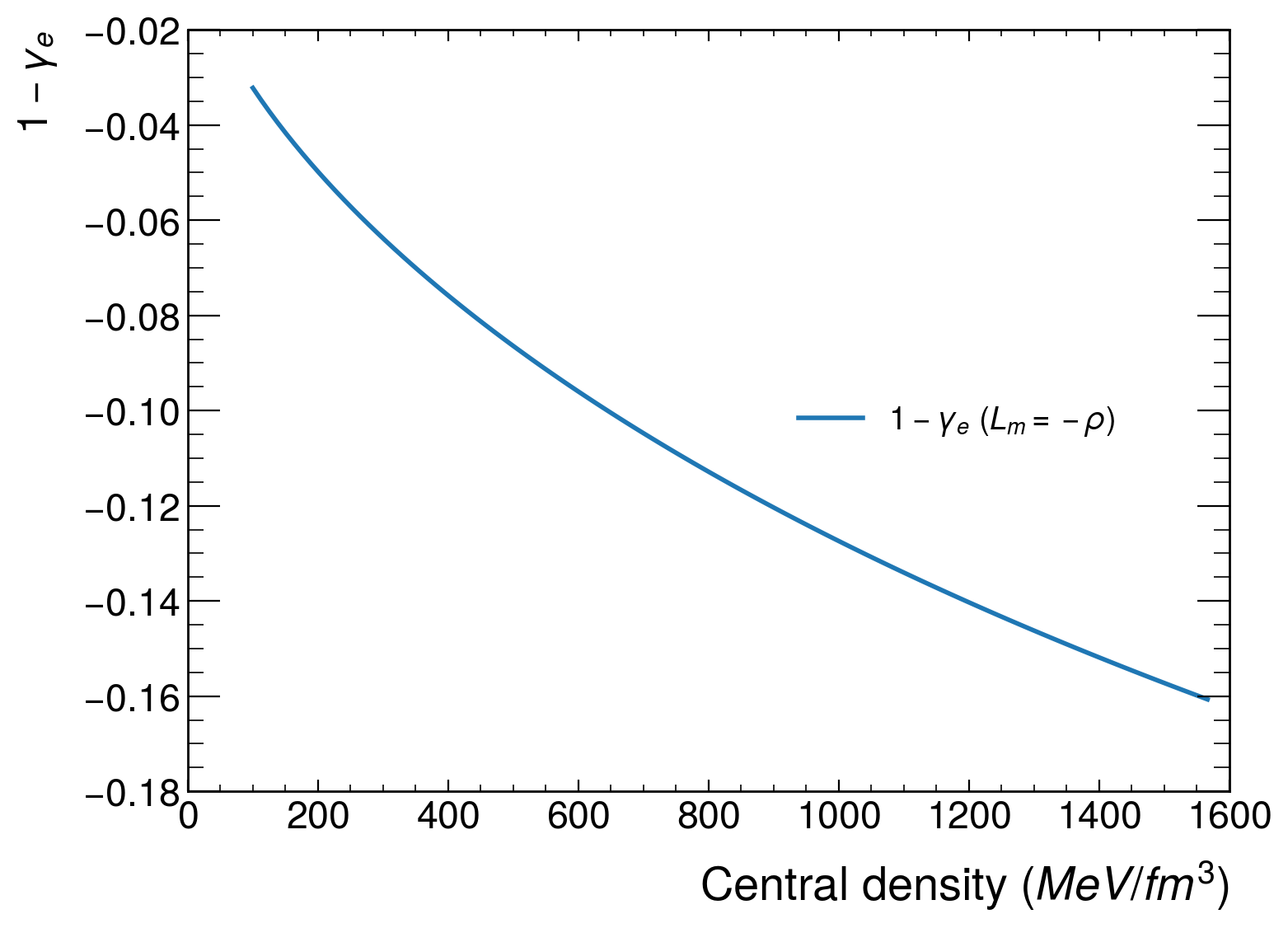}
\caption{Plot of $1-\gammae$ versus the central density of a Neutron Star. The values of central density ranges from $100$ MeV/fm$^3$ to 1578 MeV/fm$^3$. This limit comes from the fact that we want the speed of sound to be less that the conservative limit considered of $c/\sqrt 3$  \citep{bedaque:2015pr}.}\label{plot:gamma_alpha}
	\end{center}
\end{figure}

\begin{figure}[h]
	\begin{center}
		\includegraphics[scale=0.45]{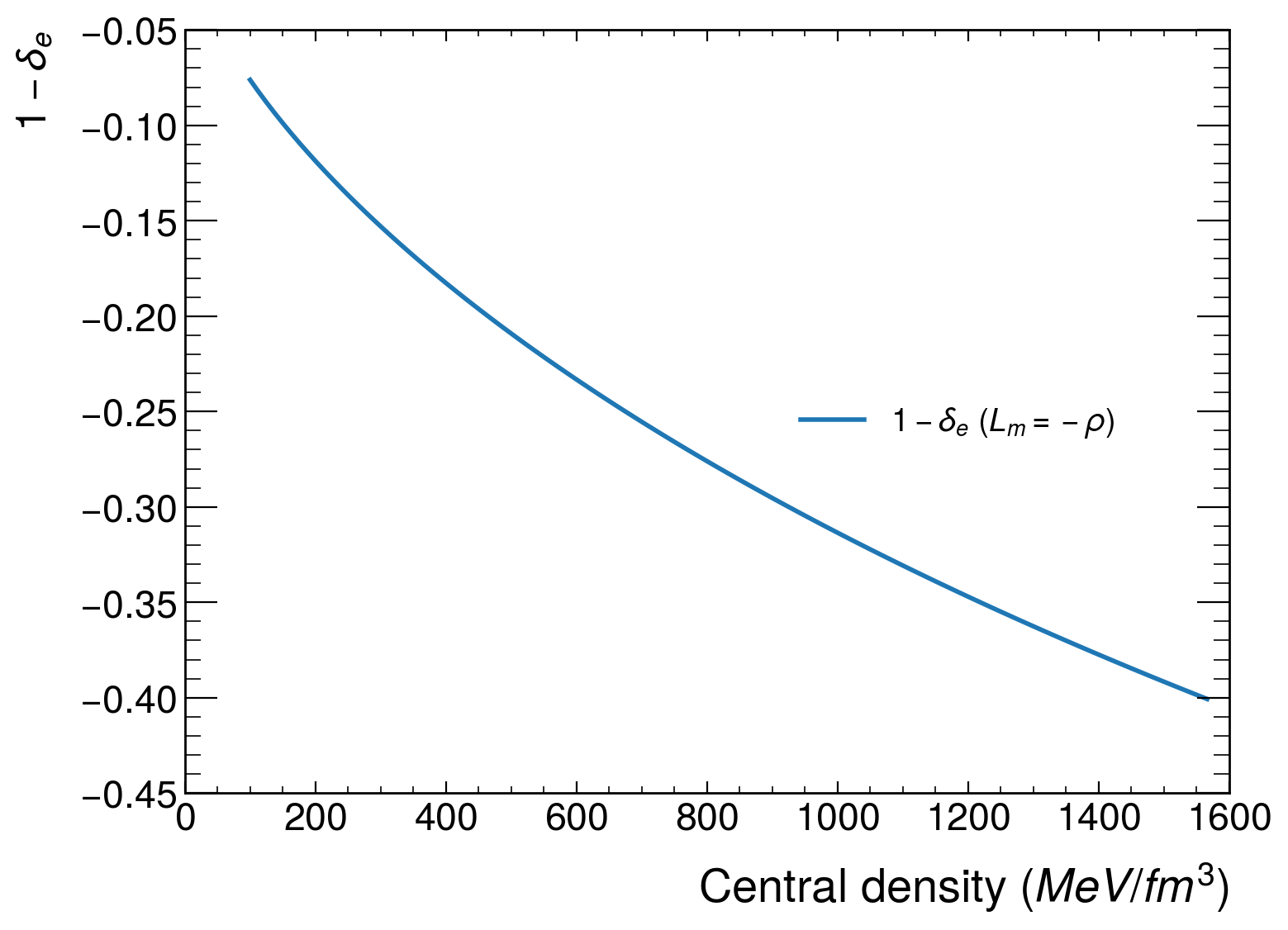}
\caption{Plot of $1-\deltae$ versus the central density of a Neutron Star. The values of central density ranges from $100$ MeV/fm$^3$ to 1578 MeV/fm$^3$. This limit comes from the fact that we want the speed of sound to be less that the conservative limit considered of $c/\sqrt 3$ \citep{bedaque:2015pr}.}\label{plot:delta_ER}
	\end{center}
\end{figure}

\section{Using dipolar gravitational waves to constrain Entangled Relativity }\label{sec:GW}

Given the large values of $1-\gammae$ predicted by Entangled Relativity for neutron stars when the on-shell matter Lagrangian is assumed to be $\Lm=-\rho$—see Fig.~\ref{plot:gamma_alpha}—it seems likely that Entangled Relativity could be heavily constrained under this assumption from observations of binary pulsars \cite{kramer:2021px}, white dwarf--pulsar systems \cite{freire:2012mn}, or triple star systems \cite{voisin:2020aa}. Indeed, one would first expect a significant deviation of the Shapiro delay, which is taken into account when fitting the system parameters to pulsar timing data \cite{kramer:2021px}. More importantly, the non-negligible numerical value of $1-\gammae$ may imply a non-negligible emission of dipolar gravitational radiation in white dwarf--pulsar systems, which is already very tightly constrained \cite{freire:2012mn,kramer:2021px}.\\

The expectation of significant dipolar emission stems from the fact that such emissions are generically proportional to the square of the Nordtvedt parameter $\eta_{PN} = 4\beta_{PN} - 3\gamma_{PN} - 1$ divided by $1-\gamma_{PN}$ \cite{freire:2012mn}, which vanishes in General Relativity. In Entangled Relativity, one has $\gamma_{PN} = \beta_{PN} = 1$—see Sec.~\ref{sec:pN}—so that $\eta_{PN}^2/(1-\gamma_{PN}) = 0$ as in General Relativity, and it would therefore seem legitimate not to expect dipolar emission. However, it is likely that dipolar emission from compact objects is sensitive to $\gammae$ and $\betae$, rather than to $\gamma_{PN}$ and $\beta_{PN}$. Hence, one may still expect dipolar emission in white dwarf--pulsar systems in particular.

Using the correspondence between the exact post-Newtonian parameters and the Damour-Esposito-Farèse scalar charge---described in Eq. (\ref{eq:alphaDEF-alphaJNW}) and which remains valid in Entangled Relativity---we will now see how binary compact objects radiate dipolar gravitational waves.
In this part we will show how to evaluate the variation in orbital period predicted by Entangled Relativity for a binary system composed of a pulsar and a white dwarf emitting dipolar gravitational waves. To accomplish this, we again use our Tolmann-Oppenheimer-Volkoff simulation in Entangled Relativity \citep{arruga:2021pr}, with the same polytropic equation of state, as well as with another, more realistic, equation of state. Then we use the analytical external solution to deduce the scalar charge as explained in Sec.~\ref{sec:verification}, as this one intervenes directly in the computation of the variation of the orbital period.

\subsection{Evaluating the dipolar radiation predicted by Entangled Relativity}

With the method in Sec. \ref{sec:DEFvsGammae}, one can verify that Eq. (\ref{eq:alphaDEF-alphaJNW}) also applies to Entangled Relativity, provided that one imposes $\omega_0 = 0$.
Following \cite{freire:2012mn}, the time variation of the dipolar period $\dot{P}^D$ for a binary system in scalar-tensor theory is given by
\be \label{eq:Pdot}
\dot{P}^D = - 2 \pi n_b \mu_* c^{-3} \frac{q}{q+1} \frac{1+e^2/2}{\left(1-e^2\right)^{5/2}} \left( \alpha_P - \alpha_c \right)^2,
\ee
where $n_b$ is the orbital frequency, $\mu_*$ is the gravitational parameter, $q$ is the mass ratio between the pulsar and the companion, $e$ is the eccentricity and $\alpha_X \equiv \alpha_{DEF}$ is the Damour-Esposito-Farèse scalar charge of the body $X$ where $P$ stands for \textit{Pulsar} and $c$ for \textit{companion}.\\

As the parameters $\gammae$ and $\alpha_X$ are linked by Eq. (\ref{eq:alphaDEF-alphaJNW}) one can rewrite the orbital decay rate Eq. (\ref{eq:Pdot}) as 


\be
\begin{aligned}
\label{eq:Pdot_exact}
     \dot{P}^D = - 24 \pi n_b& \mu_* c^{-3} \frac{q}{q+1} \frac{1+e^2/2}{\left(1-e^2\right)^{5/2}}\\ 
     \times&\left( \frac{\left( \gamma_{ec} - \gamma_{eP}\right)^2}{\left( 1+ \gamma_{ec} \right)^2 \left( 1+ \gamma_{eP} \right)^2} \right),
\end{aligned}
\ee
where $\gamma_{eP}$ and $\gamma_{ec}$ are respectively the exact Post-Newtonian parameter $\gammae$ for the neutron star and for the companion.

We focus on the relativistic Pulsar-White dwarf low eccentricity binary system PSR J1738+0333.\footnote{In the sense that we use the inferred parameters values of this system in General Relativity to estimate the amount of dipolar gravitational radiation that one could expect for this type of systems. However, one has to keep in mind that all the parameters are model dependent and have to be readjusted within the new theoretical framework.} These types of binary systems are extremely fruitful for testing alternative theories of gravity. Indeed, they predict a differential in scalar charge that leads to the emission of dipolar gravitational waves. Since the effect of the latter has not been observed yet, these studies can put stringent constraints on scalar-tensor theories of gravity.

\begin{table}[ht]
\centering
\begin{tabular}{|c|c|}\hline
\makebox[6em]{Parameter}&\makebox[6em]{Value}\\\hline\hline
$n_b$ & $2.05 \times 10^{-4} s^{-1}$\\
\hline
$\mu_*$&$G*M_c = 2.40\times10^{19}m^3 s^{-2}$\\
\hline
M$_p$&computed\\
\hline
$q$ & 8.1 \\
\hline
$e$ & $ 3.4\times 10^{-7}$ \\
\hline
$\alpha_p$ & computed \\
\hline
$\alpha_c$ & 0\footnote{Using the TOV framework applied to the case of white dwarfs \citep{chehab:2025cq} shows that $\alpha_c \in [-2.72\times10^{-5},1.21\times10^{-3}]$. Using these values instead of $0$ for $\alpha_c$ leads to negligible corrections to our estimates.} \\
\hline
\end{tabular}
\caption{Summary of the parameters used in order to compute $\dot{P}^D$ for the system PSR J1738+0333. The parameters come from \cite{freire:2012mn}. Here, $G_*$ is the value of Newton's constant measured thanks to a Cavendish experiment. The pulsar's mass and its scalar charge $\alpha_p$ have been computed in the case of Entangled Relativity while $\alpha_c$ is negligible.} \label{Table:value}
\end{table}

To numerically evaluate the variation of the orbital period, we use our Tolman-Oppenheimer-Volkoff simulation presented before (see Sec. \ref{sec:TOV} or \cite{arruga:2021pr}). But because it assumes a too simple polytropic equation of state, we also run the simulation with a more realistic, piecewise, Skyrme Lyon (SLy) equation of state \cite{Potekhin2004:AA,Douchin2001:AA}, which has been coded in \citep{deniscode}.\footnote{The adapted code to compute $\dot{P}$ from this piece wise equation of state is given in \citep{thomascode}.} Using the values presented in Table \ref{Table:value}, we ran the simulation for $300$ central densities between $100$MeV/fm$^3$ and $2000$ MeV/fm$^3$ but we only kept simulations for which the speed of sound is less than the conservative limit usually considered of $c/\sqrt{3}$ \citep{bedaque:2015pr} in the whole neutron star.} However, we also limit our analysis to neutron star masses greater than one solar mass. \\

\begin{figure}[h]
	\begin{center}
		\includegraphics[scale=0.45]{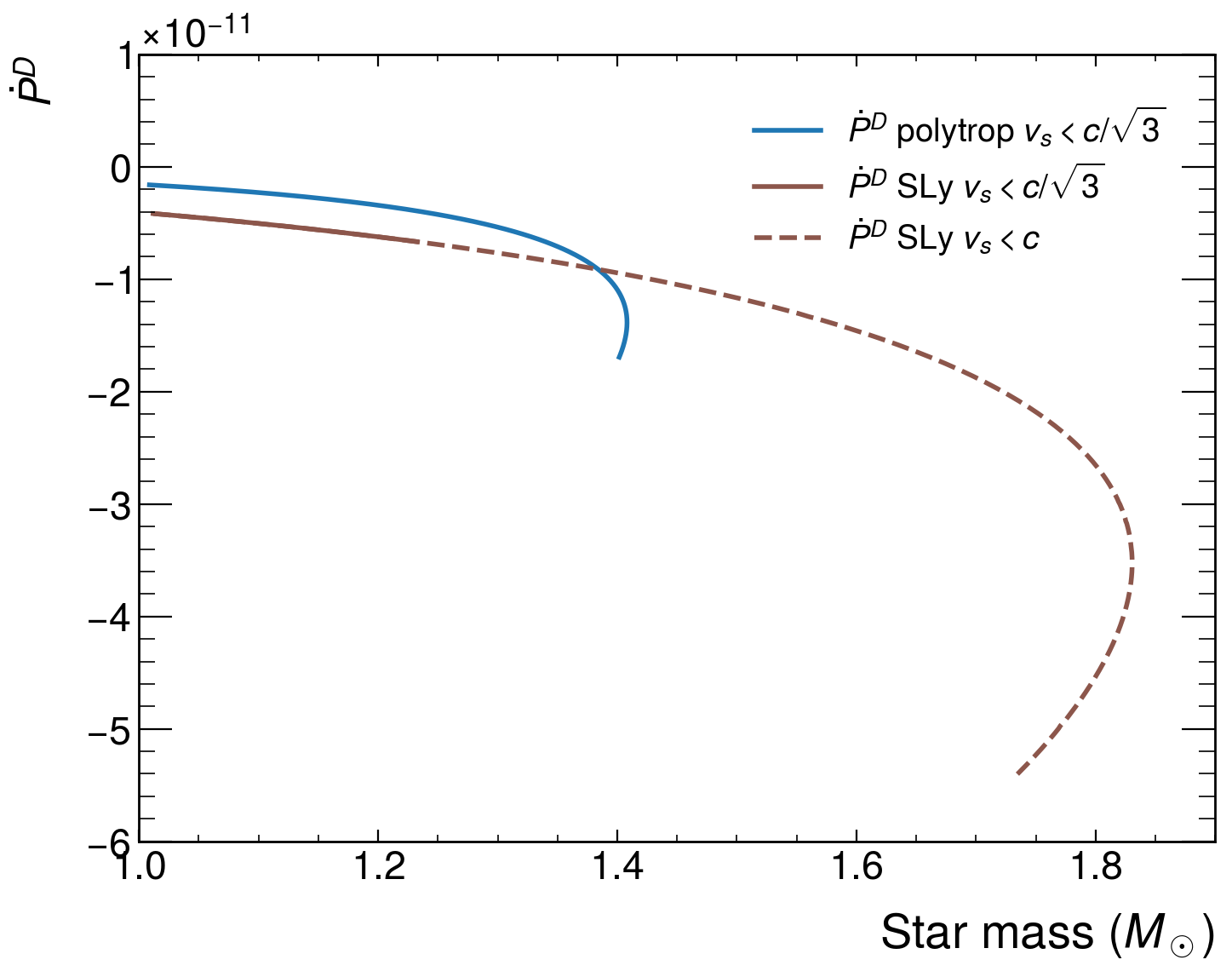}
\caption{Plot of $\dot P^D$ versus the predicted masses of the Neutron Stars for two different equation of state.}\label{plot:pdot_ER}
	\end{center}
\end{figure}

The values obtained from our simulation in Entangled Relativity range from $\dot{P}^D = -1.69\times 10^{-11} ss^{-1}$ for a mass of $1.40M_\odot$ to $\dot{P}^D = -1.62\times 10^{-12} ss^{-1}$ for a mass of $1.01M_\odot$ for a polytropic equation of state, and  $\dot{P}^D = -6.67\times 10^{-12} ss^{-1}$ for a mass of $1.23M_\odot$ to $\dot{P}^D = -4.17\times 10^{-12} ss^{-1}$ for a mass of $1.01M_\odot$ for a piece wise equation of state. Our low mass estimation comes from the speed-of-sound limit of $c/\sqrt{3}$. Indeed, this limit appears to impose constraints on the masses of the simulated neutron stars. In particular, there is a significant tension---in General Relativity already---between this velocity bound and neutron stars with masses around two solar mass \citep{bedaque:2015pr}. The details of our simulations are presented in Fig. \ref{plot:pdot_ER}, where we also present the case where one relaxes the speed of sound limit for the piece wise equation of state in order to illustrate how the speed of sound limit impacts the value of the maximal mass attainable. \\

On the other hand, the data analysis in \cite{freire:2012mn} imply that one has
\be \label{eq:obs_pdot}
\dot{P} =\left( -25.9 \pm 3.2 \right)10^{-15} s s^{-1}.
\ee
Comparing our results presented above and in Fig. \ref{plot:pdot_ER} to the inferred value Eq. (\ref{eq:obs_pdot}) seems to indicate that the case $\Lm = - \rho$ should be heavily constrained, if not ruled out, by this type of systems. But in order to evaluate precisely to what extent, one would have to readjust all the parameters of the binary system within the framework of Entangled Relativity and to also check a wide range of possible equations of state.

\subsection{Discussion}

At best, our simulations show that the predictions differ by two orders of magnitude in Entangled Relativity with the assumption that $\Lm=-\rho$ (see Fig. \ref{plot:pdot_ER}) and the inferred constraint from observational data in Eq. (\ref{eq:obs_pdot}). This seems to indicate that the case $\Lm = - \rho$ should be heavily constrained, if not ruled out, by such systems. But in order to evaluate precisely to what extent, one would have to readjust all the parameters within the framework of Entangled Relativity and to also check a wide range of possible equations of state.
Still, it is unlikely that another set of parameters and equations of state will be able to recover the observed orbital period data---given the orders of magnitude apparent difference.

Even if this result is confirmed, Entangled Relativity would remain viable if $\Lm = T$ instead, as it has sometimes been argued in the literature \citep{avelino:2018pr,avelino:2022pr,pinto:2025ar}. But then, Entangled Relativity would not predict any deviation from General Relativity for compact objects that are not extremely magnetized. Indeed, in that situation, only the magnetic field would source the scalar field.
Hence a deviation from General Relativity should be expected for magnetars, but the source term of the scalar field is $B^2/2\mu_0$ and is around $10^{11}$ Tesla \citep{vidana2018}), which is typically $10^{-7}$ relatively smaller than the typical energy densities of neutron stars---or their pressure, since $\rho\sim 3 P/c^2$ for the densest neutron stars. Therefore, a deviation from General Relativity that would entirely be due to a magnetic field with the assumption $\Lm=T$ is expected to be up to seven orders of magnitude smaller than one expects from the matter density with the assumption $\Lm=-\rho$, see \citep{chehab:2025cq}.

\section{Conclusion}
\label{sec:concl}

In this work, and for the first time, we have numerically evaluated the so-called \emph{exact} post-Newtonian parameters in scalar--tensor theories of gravity, focusing on Brans--Dicke theory and on Entangled Relativity. Contrary to the usual post-Newtonian parameters, which are perturbative quantities defined in the weak-field limit, the exact parameters are non-perturbative and explicitly depend on the internal structure of the gravitating body through its energy density and pressure. As such, they remain meaningful in the strong-field regime of compact objects, where the standard post-Newtonian expansion becomes insufficient.\\

We have developed two independent numerical methods to compute these exact parameters: a direct evaluation based on Tolman--Oppenheimer--Volkoff solutions, and an indirect determination through the matching of the external spacetime to the Janis--Newman--Winicour solution. The excellent agreement between these two approaches provides a strong validation of both the numerical implementation and the analytical expressions previously derived in \cite{chauvineau:2024plb,chauvineau:2024prd}.\\

In the case of Brans--Dicke theory, we have shown that the exact parameters can differ significantly from their usual post-Newtonian counterparts, with relative deviations reaching several tens of percent for sufficiently compact objects. Interestingly, we found that for ultra-relativistic stars the sourcing of the scalar field is reduced due to the relation between pressure and energy density, leading in some cases to smaller deviations from General Relativity than might have been naively expected. These results highlight that strong-field effects cannot be reliably inferred from weak-field post-Newtonian parameters alone.\\

We then applied the same framework to Entangled Relativity. While the usual post-Newtonian analysis of this theory yields $\gamma_{PN}=\beta_{PN}=1$, we have shown that this result does not extend to the exact parameters in the strong-field regime. For neutron stars, and under the assumption that the on-shell matter Lagrangian is $\Lm=-\rho$, the deviations of $\gammae$ and $\deltae$ from unity can become substantial, reaching the level of several percent or more, implying indirectly the prediction of dipolar gravitational waves in Entangled Relativity.

In order to probe this different prediction with respect to General Relativity, we showed that the variation in orbital period in Entangled Relativity when assuming $\Lm = - \rho$ ranges from two to three orders of magnitude above the constraint inferred from observational data. This suggests that the theory may already be ruled out by existing data, provided that the assumption $\mathcal{L}_m = -\rho$ holds; however, confirming this requires constraining the binary system’s parameters with observational data under various possible equations of state.

If one assumes $\Lm=T$ instead, as advocated in several works, the scalar field is not sourced by ordinary matter and the theory reduces more effectively to General Relativity for compact objects. In that case, observable deviations would only arise for objects endowed with extremely strong electromagnetic fields, such as magnetars, and are in any case expected to be several orders of magnitude smaller than what is found with the assumption $\Lm=-\rho$.

\section*{Acknowledgement}
The authors would like to thank Bertrand Chauvineau for his insightful comments, for suggesting ways to identify the exact moment at which the pressure vanished numerically, and ways for checking the validity of our integration.

\bibliography{GammaEval}

@ARTICLE{minazzoli:2026pl,
       author = {{Minazzoli}, Olivier and {Wavasseur}, Maxime and {Chehab}, Thomas},
        title = "{Deriving entangled relativity}",
      journal = {Physics Letters B},
         year = 2026,
        month = feb,
       volume = {873},
          eid = {140117},
        pages = {140117},
          doi = {10.1016/j.physletb.2025.140117},
archivePrefix = {arXiv},
       eprint = {2506.15209},
 primaryClass = {gr-qc},
       adsurl = {https://ui.adsabs.harvard.edu/abs/2026PhLB..87340117M}
}

@ARTICLE{minazzoli:2011cq,
       author = {{Minazzoli}, Olivier and {Chauvineau}, Bertrand},
        title = "{Scalar-tensor propagation of light in the inner solar system including relevant c$^{-4}$ contributions for ranging and time transfer}",
      journal = {Classical and Quantum Gravity},
         year = 2011,
        month = apr,
       volume = {28},
       number = {8},
          eid = {085010},
        pages = {085010},
          doi = {10.1088/0264-9381/28/8/085010},
archivePrefix = {arXiv},
       eprint = {1007.3942},
 primaryClass = {gr-qc},
       adsurl = {https://ui.adsabs.harvard.edu/abs/2011CQGra..28h5010M}
}

@ARTICLE{ashby:2010cq,
       author = {{Ashby}, Neil and {Bertotti}, Bruno},
        title = "{Accurate light-time correction due to a gravitating mass}",
      journal = {Classical and Quantum Gravity},
         year = 2010,
        month = jul,
       volume = {27},
       number = {14},
          eid = {145013},
        pages = {145013},
          doi = {10.1088/0264-9381/27/14/145013},
archivePrefix = {arXiv},
       eprint = {0912.2705},
 primaryClass = {gr-qc},
       adsurl = {https://ui.adsabs.harvard.edu/abs/2010CQGra..27n5013A}
}

@article{deng:2012pr,
  title = {Two-post-Newtonian light propagation in the scalar-tensor theory: An $N$-point mass case},
  author = {Deng, Xue-Mei and Xie, Yi},
  journal = {Phys. Rev. D},
  volume = {86},
  issue = {4},
  pages = {044007},
  numpages = {17},
  year = {2012},
  month = {Aug},
  publisher = {American Physical Society},
  doi = {10.1103/PhysRevD.86.044007},
  url = {https://link.aps.org/doi/10.1103/PhysRevD.86.044007}
}

@ARTICLE{linet:2013cq,
       author = {{Linet}, Bernard and {Teyssandier}, Pierre},
        title = "{New method for determining the light travel time in static, spherically symmetric spacetimes. Calculation of the terms of order G$^{3}$}",
      journal = {Classical and Quantum Gravity},
         year = 2013,
        month = sep,
       volume = {30},
       number = {17},
          eid = {175008},
        pages = {175008},
          doi = {10.1088/0264-9381/30/17/175008},
archivePrefix = {arXiv},
       eprint = {1304.3683},
 primaryClass = {gr-qc},
       adsurl = {https://ui.adsabs.harvard.edu/abs/2013CQGra..30q5008L}
}

@ARTICLE{hees:2014pr,
       author = {{Hees}, A. and {Bertone}, S. and {Le Poncin-Lafitte}, C.},
        title = "{Relativistic formulation of coordinate light time, Doppler, and astrometric observables up to the second post-Minkowskian order}",
      journal = {\prd},
         year = 2014,
        month = mar,
       volume = {89},
       number = {6},
          eid = {064045},
        pages = {064045},
          doi = {10.1103/PhysRevD.89.064045},
archivePrefix = {arXiv},
       eprint = {1401.7622},
 primaryClass = {gr-qc},
       adsurl = {https://ui.adsabs.harvard.edu/abs/2014PhRvD..89f4045H}
}

@ARTICLE{linet:2016pr,
       author = {{Linet}, B. and {Teyssandier}, P.},
        title = "{Time transfer functions in Schwarzschild-like metrics in the weak-field limit: A unified description of Shapiro and lensing effects}",
      journal = {\prd},
         year = 2016,
        month = feb,
       volume = {93},
       number = {4},
          eid = {044028},
        pages = {044028},
          doi = {10.1103/PhysRevD.93.044028},
archivePrefix = {arXiv},
       eprint = {1511.04284},
 primaryClass = {gr-qc},
       adsurl = {https://ui.adsabs.harvard.edu/abs/2016PhRvD..93d4028L}
}

@ARTICLE{cappuccio:2021cq,
       author = {{Cappuccio}, Paolo and {di Stefano}, Ivan and {Cascioli}, Gael and {Iess}, Luciano},
        title = "{Comparison of light-time formulations in the post-Newtonian framework for the BepiColombo MORE experiment}",
      journal = {Classical and Quantum Gravity},
         year = 2021,
        month = nov,
       volume = {38},
       number = {22},
          eid = {227001},
        pages = {227001},
          doi = {10.1088/1361-6382/ac2b0a},
archivePrefix = {arXiv},
       eprint = {2201.05092},
 primaryClass = {gr-qc},
       adsurl = {https://ui.adsabs.harvard.edu/abs/2021CQGra..38v7001C}
}

@ARTICLE{zschocke:2022pr,
       author = {{Zschocke}, Sven},
        title = "{Light propagation in 2PN approximation in the monopole and quadrupole field of a body at rest: Initial value problem}",
      journal = {\prd},
         year = 2022,
        month = jan,
       volume = {105},
       number = {2},
          eid = {024040},
        pages = {024040},
          doi = {10.1103/PhysRevD.105.024040},
archivePrefix = {arXiv},
       eprint = {2201.06296},
 primaryClass = {gr-qc},
       adsurl = {https://ui.adsabs.harvard.edu/abs/2022PhRvD.105b4040Z}
}

@article{minazzoli:2025pl,
    author = "Minazzoli, Olivier and Wavasseur, Maxime and Chehab, Thomas",
    title = "{Deriving entangled relativity}",
    eprint = "2506.15209",
    archivePrefix = "arXiv",
    primaryClass = "gr-qc",
    doi = "10.1016/j.physletb.2025.140117",
    journal = "Phys. Lett. B",
    volume = "873",
    pages = "140117",
    year = "2026"
}

@ARTICLE{damour:1996pr,
       author = {{Damour}, Thibault and {Esposito-Far{\`e}se}, Gilles},
        title = "{Tensor-scalar gravity and binary-pulsar experiments}",
      journal = {\prd},
         year = 1996,
        month = jul,
       volume = {54},
       number = {2},
        pages = {1474-1491},
          doi = {10.1103/PhysRevD.54.1474},
archivePrefix = {arXiv},
       eprint = {gr-qc/9602056},
 primaryClass = {gr-qc},
       adsurl = {https://ui.adsabs.harvard.edu/abs/1996PhRvD..54.1474D}
}

@ARTICLE{damour:1993pl,
       author = {{Damour}, Thibault and {Esposito-Farese}, Gilles},
        title = "{Nonperturbative strong-field effects in tensor-scalar theories of gravitation}",
      journal = {\prl},
         year = 1993,
        month = apr,
       volume = {70},
       number = {15},
        pages = {2220-2223},
          doi = {10.1103/PhysRevLett.70.2220},
       adsurl = {https://ui.adsabs.harvard.edu/abs/1993PhRvL..70.2220D}
}

@misc{vidana2018,
      title={A short walk through the physics of neutron stars}, 
      author={Isaac Vidana},
      year={2018},
      eprint={1805.00837},
      archivePrefix={arXiv},
      primaryClass={nucl-th},
      url={https://arxiv.org/abs/1805.00837}, 
}

@article{kramer:2021px,
  title = {Strong-Field Gravity Tests with the Double Pulsar},
  author = {Kramer, M. and Stairs, I. H. and Manchester, R. N. and Wex, N. and Deller, A. T. and Coles, W. A. and Ali, M. and Burgay, M. and Camilo, F. and Cognard, I. and Damour, T. and Desvignes, G. and Ferdman, R. D. and Freire, P. C. C. and Grondin, S. and Guillemot, L. and Hobbs, G. B. and Janssen, G. and Karuppusamy, R. and Lorimer, D. R. and Lyne, A. G. and McKee, J. W. and McLaughlin, M. and M\"unch, L. E. and Perera, B. B. P. and Pol, N. and Possenti, A. and Sarkissian, J. and Stappers, B. W. and Theureau, G.},
  journal = {Phys. Rev. X},
  volume = {11},
  issue = {4},
  pages = {041050},
  numpages = {53},
  year = {2021},
  month = {Dec},
  publisher = {American Physical Society},
  doi = {10.1103/PhysRevX.11.041050},
  url = {https://link.aps.org/doi/10.1103/PhysRevX.11.041050}
}

@ARTICLE{freire:2012mn,
       author = {{Freire}, Paulo C.~C. and {Wex}, Norbert and {Esposito-Far{\`e}se}, Gilles and {Verbiest}, Joris P.~W. and {Bailes}, Matthew and {Jacoby}, Bryan A. and {Kramer}, Michael and {Stairs}, Ingrid H. and {Antoniadis}, John and {Janssen}, Gemma H.},
        title = "{The relativistic pulsar-white dwarf binary PSR J1738+0333 - II. The most stringent test of scalar-tensor gravity}",
      journal = {\mnras},
         year = 2012,
        month = jul,
       volume = {423},
       number = {4},
        pages = {3328-3343},
          doi = {10.1111/j.1365-2966.2012.21253.x},
archivePrefix = {arXiv},
       eprint = {1205.1450},
 primaryClass = {astro-ph.GA},
       adsurl = {https://ui.adsabs.harvard.edu/abs/2012MNRAS.423.3328F}
}

@article{chehab:2025cq,
doi = {10.1088/1361-6382/ae30c7},
url = {https://doi.org/10.1088/1361-6382/ae30c7},
year = {2025},
month = {dec},
publisher = {IOP Publishing},
author = {{Chehab}, T. and {Minazzoli}, O. and {Hees}, A.},
title = {Variation of Planck’s quantum of action in Entangled Relativity},
journal = {Classical and Quantum Gravity}
}

@ARTICLE{voisin:2020aa,
       author = {{Voisin}, G. and {Cognard}, I. and {Freire}, P.~C.~C. and {Wex}, N. and {Guillemot}, L. and {Desvignes}, G. and {Kramer}, M. and {Theureau}, G.},
        title = "{An improved test of the strong equivalence principle with the pulsar in a triple star system}",
      journal = {\aap},
         year = 2020,
        month = jun,
       volume = {638},
          eid = {A24},
        pages = {A24},
          doi = {10.1051/0004-6361/202038104},
archivePrefix = {arXiv},
       eprint = {2005.01388},
 primaryClass = {gr-qc},
       adsurl = {https://ui.adsabs.harvard.edu/abs/2020A&A...638A..24V}
}

@ARTICLE{EHT:2025lr,
       author = {{Ayzenberg}, Dimitry and {Blackburn}, Lindy and {Brito}, Richard and {Britzen}, Silke and {Broderick}, Avery E. and {Carballo-Rubio}, Ra{\'u}l and {Cardoso}, Vitor and {Chael}, Andrew and {Chatterjee}, Koushik and {Chen}, Yifan and {Cunha}, Pedro V.~P. and {Davoudiasl}, Hooman and {Denton}, Peter B. and {Doeleman}, Sheperd S. and {Eichhorn}, Astrid and {Eubanks}, Marshall and {Fang}, Yun and {Foschi}, Arianna and {Fromm}, Christian M. and {Galison}, Peter and {Ghosh}, Sushant G. and {Gold}, Roman and {Gurvits}, Leonid I. and {Hadar}, Shahar and {Held}, Aaron and {Houston}, Janice and {Hu}, Yichao and {Johnson}, Michael D. and {Kocherlakota}, Prashant and {Natarajan}, Priyamvada and {Olivares}, H{\'e}ctor and {Palumbo}, Daniel and {Pesce}, Dominic W. and {Rajendran}, Surjeet and {Roy}, Rittick and {Saurabh} and {Shao}, Lijing and {Tahura}, Shammi and {Tamar}, Aditya and {Tiede}, Paul and {Vincent}, Fr{\'e}d{\'e}ric H. and {Visinelli}, Luca and {Wang}, Zhiren and {Wielgus}, Maciek and {Xue}, Xiao and {Yakut}, Kadri and {Yang}, Huan and {Younsi}, Ziri},
        title = "{Author Correction: Fundamental physics opportunities with the next-generation Event Horizon Telescope}",
      journal = {Living Reviews in Relativity},
         year = 2025,
        month = sep,
       volume = {28},
       number = {1},
          eid = {7},
        pages = {7},
          doi = {10.1007/s41114-025-00062-3},
       adsurl = {https://ui.adsabs.harvard.edu/abs/2025LRR....28....7A}
}

@ARTICLE{williams:2009ij,
       author = {{Williams}, James G. and {Turyshev}, Slava G. and {Boggs}, Dale H.},
        title = "{Lunar Laser Ranging Tests of the Equivalence Principle with the Earth and Moon}",
      journal = {International Journal of Modern Physics D},
         year = 2009,
        month = jan,
       volume = {18},
       number = {7},
        pages = {1129-1175},
          doi = {10.1142/S021827180901500X},
archivePrefix = {arXiv},
       eprint = {gr-qc/0507083},
 primaryClass = {gr-qc},
       adsurl = {https://ui.adsabs.harvard.edu/abs/2009IJMPD..18.1129W}
}

@article{Minazzoli_2018,
   title={Rethinking the link between matter and geometry},
   volume={98},
   ISSN={2470-0029},
   url={http://dx.doi.org/10.1103/PhysRevD.98.124020},
   DOI={10.1103/physrevd.98.124020},
   number={12},
   journal={Physical Review D},
   publisher={American Physical Society (APS)},
   author={Minazzoli, Olivier},
   year={2018},
   month=dec }

@article{janis:1968prl,
  title = {Reality of the Schwarzschild Singularity},
  author = {Janis, Allen I. and Newman, Ezra T. and Winicour, Jeffrey},
  journal = {Phys. Rev. Lett.},
  volume = {20},
  issue = {16},
  pages = {878--880},
  numpages = {0},
  year = {1968},
  month = {Apr},
  publisher = {American Physical Society},
  doi = {10.1103/PhysRevLett.20.878},
  url = {https://link.aps.org/doi/10.1103/PhysRevLett.20.878}
}

@article{T_Damour_1992,
doi = {10.1088/0264-9381/9/9/015},
url = {https://dx.doi.org/10.1088/0264-9381/9/9/015},
year = {1992},
month = {sep},
publisher = {},
volume = {9},
number = {9},
pages = {2093},
author = {T Damour and G Esposito-Farese},
title = {Tensor-multi-scalar theories of gravitation},
journal = {Classical and Quantum Gravity}
}

@article{minazzoli:2025ej,
    author = "Minazzoli, Olivier and Wavasseur, Maxime",
    title = "{Compact objects with scalar charge embedded in a magnetic or electric field in Einstein\textendash{}Maxwell-dilaton theories}",
    eprint = "2502.13829",
    archivePrefix = "arXiv",
    primaryClass = "gr-qc",
    doi = "10.1140/epjc/s10052-025-14179-w",
    journal = "Eur. Phys. J. C",
    volume = "85",
    number = "4",
    pages = "474",
    year = "2025"
}

@ARTICLE{abbott:2020lr,
       author = {LVK collaboration},
        title = "{Prospects for observing and localizing gravitational-wave transients with Advanced LIGO, Advanced Virgo and KAGRA}",
      journal = {Living Reviews in Relativity},
         year = 2020,
        month = dec,
       volume = {23},
       number = {1},
          eid = {3},
        pages = {3},
          doi = {10.1007/s41114-020-00026-9},
       adsurl = {https://ui.adsabs.harvard.edu/abs/2020LRR....23....3A}
}

@ARTICLE{fienga:2024lr,
       author = {{Fienga}, Agn{\`e}s and {Minazzoli}, Olivier},
        title = "{Testing theories of gravity with planetary ephemerides}",
      journal = {Living Reviews in Relativity},
         year = 2024,
        month = dec,
       volume = {27},
       number = {1},
          eid = {1},
        pages = {1},
          doi = {10.1007/s41114-023-00047-0},
archivePrefix = {arXiv},
       eprint = {2303.01821},
 primaryClass = {gr-qc},
       adsurl = {https://ui.adsabs.harvard.edu/abs/2024LRR....27....1F}
}

@ARTICLE{pinto:2025ar,
       author = {{Pinto}, S.~R. and {Avelino}, P.~P.},
        title = "{Deviations from the von Laue Condition: Implications for the On-Shell Lagrangian of Particles and Fluids}",
      journal = {arXiv e-prints},
         year = 2025,
        month = feb,
          eid = {arXiv:2502.10427},
        pages = {arXiv:2502.10427},
archivePrefix = {arXiv},
       eprint = {2502.10427},
 primaryClass = {gr-qc},
       adsurl = {https://ui.adsabs.harvard.edu/abs/2025arXiv250210427P}
}

@ARTICLE{minazzoli:2024pn,
       author = {{Minazzoli}, O.},
        title = "{On the Principle of Relativity of Inertia in Both General and Entangled Relativities}",
      journal = {Physics of Particles and Nuclei},
         year = 2024,
        month = dec,
       volume = {55},
       number = {6},
        pages = {1488-1493},
          doi = {10.1134/S1063779624701132},
       adsurl = {https://ui.adsabs.harvard.edu/abs/2024PPN....55.1488M}
}

@article{wavasseur:2025gg,
   title={Slowly rotating and charged black-holes in entangled relativity},
   volume={57},
   ISSN={1572-9532},
   url={http://dx.doi.org/10.1007/s10714-025-03366-5},
   DOI={10.1007/s10714-025-03366-5},
   number={2},
   journal={General Relativity and Gravitation},
   publisher={Springer Science and Business Media LLC},
   author={Wavasseur, Maxime and Abrial, Théo and Minazzoli, Olivier},
   year={2025},
    eprint={2411.09327},
    archivePrefix={arXiv},
    primaryClass={gr-qc},
   month=feb }

@ARTICLE{damour:1992cq,
       author = {{Damour}, T. and {Esposito-Farese}, G.},
        title = "{Tensor-multi-scalar theories of gravitation}",
      journal = {Classical and Quantum Gravity},
         year = 1992,
        month = sep,
       volume = {9},
       number = {9},
        pages = {2093-2176},
          doi = {10.1088/0264-9381/9/9/015},
       adsurl = {https://ui.adsabs.harvard.edu/abs/1992CQGra...9.2093D}
}

@ARTICLE{garfinkle:1991pr,
       author = {{Garfinkle}, David and {Horowitz}, Gary T. and {Strominger}, Andrew},
        title = "{Charged black holes in string theory}",
      journal = {\prd},
         year = 1991,
        month = may,
       volume = {43},
       number = {10},
        pages = {3140-3143},
          doi = {10.1103/PhysRevD.43.3140},
       adsurl = {https://ui.adsabs.harvard.edu/abs/1991PhRvD..43.3140G}
}

@ARTICLE{bedaque:2015pr,
       author = {{Bedaque}, Paulo and {Steiner}, Andrew W.},
        title = "{Sound Velocity Bound and Neutron Stars}",
      journal = {\prl},
         year = 2015,
        month = jan,
       volume = {114},
       number = {3},
          eid = {031103},
        pages = {031103},
          doi = {10.1103/PhysRevLett.114.031103},
archivePrefix = {arXiv},
       eprint = {1408.5116},
 primaryClass = {nucl-th},
       adsurl = {https://ui.adsabs.harvard.edu/abs/2015PhRvL.114c1103B}
}

@ARTICLE{avelino:2018pr,
       author = {{Avelino}, P.~P. and {Azevedo}, R.~P.~L.},
        title = "{Perfect fluid Lagrangian and its cosmological implications in theories of gravity with nonminimally coupled matter fields}",
      journal = {\prd},
         year = 2018,
        month = mar,
       volume = {97},
       number = {6},
          eid = {064018},
        pages = {064018},
          doi = {10.1103/PhysRevD.97.064018},
archivePrefix = {arXiv},
       eprint = {1802.04760},
 primaryClass = {gr-qc},
       adsurl = {https://ui.adsabs.harvard.edu/abs/2018PhRvD..97f4018A}
}

@ARTICLE{holzhey:1992nb,
       author = {{Holzhey}, Christoph F.~E. and {Wilczek}, Frank},
        title = "{Black holes as elementary particles}",
      journal = {Nuclear Physics B},
         year = 1992,
        month = aug,
       volume = {380},
       number = {3},
        pages = {447-477},
          doi = {10.1016/0550-3213(92)90254-9},
archivePrefix = {arXiv},
       eprint = {hep-th/9202014},
 primaryClass = {hep-th},
       adsurl = {https://ui.adsabs.harvard.edu/abs/1992NuPhB.380..447H}
}

@ARTICLE{damour:1991pr,
       author = {{Damour}, T. and {Soffel}, M. and {Xu}, C.},
        title = "{General-relativistic celestial mechanics. I. Method and definition of reference systems}",
      journal = {\prd},
         year = 1991,
        month = may,
       volume = {43},
       number = {10},
        pages = {3273-3307},
          doi = {10.1103/PhysRevD.43.3273},
       adsurl = {https://ui.adsabs.harvard.edu/abs/1991PhRvD..43.3273D}
}

@ARTICLE{capozziello:2015sc,
AUTHOR = {Capozziello, S.  and Laurentis, M. De},
TITLE   = {{F}({R}) theories of gravitation},
YEAR    = {2015},
JOURNAL = {Scholarpedia},
VOLUME  = {10},
NUMBER  = {2},
PAGES   = {31422},
DOI     = {10.4249/scholarpedia.31422},
NOTE    = {revision \#147843}
}

@ARTICLE{will:2014lr,
       author = {{Will}, Clifford M.},
        title = "{The Confrontation between General Relativity and Experiment}",
      journal = {Living Reviews in Relativity},
         year = 2014,
        month = dec,
       volume = {17},
       number = {1},
          eid = {4},
        pages = {4},
          doi = {10.12942/lrr-2014-4},
archivePrefix = {arXiv},
       eprint = {1403.7377},
 primaryClass = {gr-qc},
       adsurl = {https://ui.adsabs.harvard.edu/abs/2014LRR....17....4W}
}

@ARTICLE{minazzoli:2018pr,
       author = {{Minazzoli}, Olivier},
        title = "{Rethinking the link between matter and geometry}",
      journal = {\prd},
         year = 2018,
        month = dec,
       volume = {98},
       number = {12},
          eid = {124020},
        pages = {124020},
          doi = {10.1103/PhysRevD.98.124020},
archivePrefix = {arXiv},
       eprint = {1811.05845},
 primaryClass = {gr-qc},
       adsurl = {https://ui.adsabs.harvard.edu/abs/2018PhRvD..98l4020M}
}

@ARTICLE{ludwig:2015pl,
       author = {{Ludwig}, Hendrik and {Minazzoli}, Olivier and {Capozziello}, Salvatore},
        title = "{Merging matter and geometry in the same Lagrangian}",
      journal = {Physics Letters B},
         year = 2015,
        month = dec,
       volume = {751},
        pages = {576-578},
          doi = {10.1016/j.physletb.2015.11.023},
archivePrefix = {arXiv},
       eprint = {1506.03278},
 primaryClass = {gr-qc},
       adsurl = {https://ui.adsabs.harvard.edu/abs/2015PhLB..751..576L}
}

@ARTICLE{arruga:2021pr,
       author = {{Arruga}, Denis and {Rousselle}, Olivier and {Minazzoli}, Olivier},
        title = "{Compact objects in entangled relativity}",
      journal = {\prd},
         year = 2021,
        month = jan,
       volume = {103},
       number = {2},
          eid = {024034},
        pages = {024034},
          doi = {10.1103/PhysRevD.103.024034},
archivePrefix = {arXiv},
       eprint = {2011.14629},
 primaryClass = {gr-qc},
       adsurl = {https://ui.adsabs.harvard.edu/abs/2021PhRvD.103b4034A}
}

@ARTICLE{arruga:2021ep,
       author = {{Arruga}, Denis and {Minazzoli}, Olivier},
        title = "{Analytical external spherical solutions in entangled relativity}",
      journal = {European Physical Journal C},
         year = 2021,
        month = nov,
       volume = {81},
       number = {11},
          eid = {1027},
        pages = {1027},
          doi = {10.1140/epjc/s10052-021-09818-x},
archivePrefix = {arXiv},
       eprint = {2106.03426},
 primaryClass = {gr-qc},
       adsurl = {https://ui.adsabs.harvard.edu/abs/2021EPJC...81.1027A}
}

@ARTICLE{minazzoli:2021ej,
       author = {{Minazzoli}, Olivier and {Santos}, Edison},
        title = "{Charged black hole and radiating solutions in entangled relativity}",
      journal = {European Physical Journal C},
         year = 2021,
        month = jul,
       volume = {81},
       number = {7},
          eid = {640},
        pages = {640},
          doi = {10.1140/epjc/s10052-021-09441-w},
archivePrefix = {arXiv},
       eprint = {2102.10541},
 primaryClass = {gr-qc},
       adsurl = {https://ui.adsabs.harvard.edu/abs/2021EPJC...81..640M}
}

@ARTICLE{minazzoli:2022ar,
       author = {{Minazzoli}, Olivier},
        title = "{Quantum of action in entangled relativity}",
      journal = {arXiv e-prints},
         year = 2022,
        month = jun,
          eid = {arXiv:2206.03824},
        pages = {arXiv:2206.03824},
          doi = {10.48550/arXiv.2206.03824},
archivePrefix = {arXiv},
       eprint = {2206.03824},
 primaryClass = {gr-qc},
       adsurl = {https://ui.adsabs.harvard.edu/abs/2022arXiv220603824M}
}

@ARTICLE{einstein:1918an,
       author = {{Einstein}, A.},
        title = "{Prinzipielles zur allgemeinen Relativit{\"a}tstheorie}",
      journal = {Annalen der Physik},
         year = 1918,
        month = jan,
       volume = {360},
       number = {4},
        pages = {241-244},
          doi = {10.1002/andp.19183600402},
          note = {Translation available at \url{https://einsteinpapers.press.princeton.edu/vol7-trans/49}},
       adsurl = {https://ui.adsabs.harvard.edu/abs/1918AnP...360..241E}
}

@PROCEEDINGS{book_mach_principle,
        title = "{Mach's Principle: From Newton's Bucket to Quantum Gravity.}",
    booktitle = {Mach's Principle: From Newton's Bucket to Quantum Gravity},
         year = 1995,
        month = jan,
       adsurl = {https://ui.adsabs.harvard.edu/abs/1995mpfn.conf.....B},
      publisher = "{Birkh\"aser}",
      address   = "Boston University"
}

@INPROCEEDINGS{hoefer:1995cf,
       author = {{Hoefer}, C.},
        title = "{Einstein's Formulations of Mach's Principle}",
    booktitle = {Mach's Principle: From Newton's Bucket to Quantum Gravity},
         year = 1995,
       editor = {{Barbour}, Julian B. and {Pfister}, Herbert},
        month = jan,
        pages = {67},
       adsurl = {https://ui.adsabs.harvard.edu/abs/1995mpfn.conf...67H},
      publisher = "{Birkh\"aser}",
      address   = "Boston University"
}

@ARTICLE{minazzoli:2013pr,
       author = {{Minazzoli}, Olivier and {Hees}, Aur{\'e}lien},
        title = "{Intrinsic Solar System decoupling of a scalar-tensor theory with a universal coupling between the scalar field and the matter Lagrangian}",
      journal = {\prd},
         year = 2013,
        month = aug,
       volume = {88},
       number = {4},
          eid = {041504},
        pages = {041504},
          doi = {10.1103/PhysRevD.88.041504},
archivePrefix = {arXiv},
       eprint = {1308.2770},
 primaryClass = {gr-qc},
       adsurl = {https://ui.adsabs.harvard.edu/abs/2013PhRvD..88d1504M}
}

@ARTICLE{minazzoli:2014pr,
       author = {{Minazzoli}, Olivier and {Hees}, Aur{\'e}lien},
        title = "{Late-time cosmology of a scalar-tensor theory with a universal multiplicative coupling between the scalar field and the matter Lagrangian}",
      journal = {\prd},
         year = 2014,
        month = jul,
       volume = {90},
       number = {2},
          eid = {023017},
        pages = {023017},
          doi = {10.1103/PhysRevD.90.023017},
archivePrefix = {arXiv},
       eprint = {1404.4266},
 primaryClass = {gr-qc},
       adsurl = {https://ui.adsabs.harvard.edu/abs/2014PhRvD..90b3017M}
}

@ARTICLE{teyssandier:1983jm,
       author = {{Teyssandier}, P. and {Tourrenc}, Ph.},
        title = "{The Cauchy problem for the R+R$^{2}$ theories of gravity without torsion}",
      journal = {Journal of Mathematical Physics},
         year = 1983,
        month = dec,
       volume = {24},
       number = {12},
        pages = {2793-2799},
          doi = {10.1063/1.525659},
       adsurl = {https://ui.adsabs.harvard.edu/abs/1983JMP....24.2793T}
}

@ARTICLE{jakubiec:1988pr,
       author = {{Jakubiec}, Andrzej and {Kijowski}, Jerzy},
        title = "{On theories of gravitation with nonlinear Lagrangians}",
      journal = {\prd},
         year = 1988,
        month = mar,
       volume = {37},
       number = {6},
        pages = {1406-1409},
          doi = {10.1103/PhysRevD.37.1406},
       adsurl = {https://ui.adsabs.harvard.edu/abs/1988PhRvD..37.1406J}
}

@ARTICLE{harko:2013pr,
       author = {{Harko}, Tiberiu and {Lobo}, Francisco S.~N. and {Minazzoli}, Olivier},
        title = "{Extended f(R,L$_{m}$) gravity with generalized scalar field and kinetic term dependences}",
      journal = {\prd},
         year = 2013,
        month = feb,
       volume = {87},
       number = {4},
          eid = {047501},
        pages = {047501},
          doi = {10.1103/PhysRevD.87.047501},
archivePrefix = {arXiv},
       eprint = {1210.4218},
 primaryClass = {gr-qc},
       adsurl = {https://ui.adsabs.harvard.edu/abs/2013PhRvD..87d7501H}
}

@ARTICLE{minazzoli:2012pr,
       author = {{Minazzoli}, Olivier and {Harko}, Tiberiu},
        title = "{New derivation of the Lagrangian of a perfect fluid with a barotropic equation of state}",
      journal = {\prd},
         year = 2012,
        month = oct,
       volume = {86},
       number = {8},
          eid = {087502},
        pages = {087502},
          doi = {10.1103/PhysRevD.86.08750210.48550/arXiv.1209.2754},
archivePrefix = {arXiv},
       eprint = {1209.2754},
 primaryClass = {gr-qc},
       adsurl = {https://ui.adsabs.harvard.edu/abs/2012PhRvD..86h7502M}
}

@ARTICLE{avelino:2022pr,
       author = {{Avelino}, P.~P. and {Azevedo}, R.~P.~L.},
        title = "{On-shell Lagrangian of an ideal gas}",
      journal = {\prd},
         year = 2022,
        month = may,
       volume = {105},
       number = {10},
          eid = {104005},
        pages = {104005},
          doi = {10.1103/PhysRevD.105.104005},
archivePrefix = {arXiv},
       eprint = {2203.04022},
 primaryClass = {gr-qc},
       adsurl = {https://ui.adsabs.harvard.edu/abs/2022PhRvD.105j4005A}
}

@article{brans_Dicke:1961pr,
  title = {Mach's Principle and a Relativistic Theory of Gravitation},
  author = {Brans, C. and Dicke, R. H.},
  journal = {Phys. Rev.},
  volume = {124},
  issue = {3},
  pages = {925--935},
  numpages = {0},
  year = {1961},
  month = {Nov},
  publisher = {American Physical Society},
  doi = {10.1103/PhysRev.124.925},
  url = {https://link.aps.org/doi/10.1103/PhysRev.124.925}
}

@article{chauvineau:2024plb,
title = {The complete exterior spacetime of spherical Brans-Dicke stars},
journal = {Physics Letters B},
volume = {855},
pages = {138803},
year = {2024},
issn = {0370-2693},
doi = {https://doi.org/10.1016/j.physletb.2024.138803},
url = {https://www.sciencedirect.com/science/article/pii/S0370269324003617},
author = {Bertrand Chauvineau and Hoang Ky Nguyen}
}

@ARTICLE{arbanil:2013pr,
       author = {{Arba{\~n}il}, Jos{\'e} D.~V. and {Lemos}, Jos{\'e} P.~S. and {Zanchin}, Vilson T.},
        title = "{Polytropic spheres with electric charge: Compact stars, the Oppenheimer-Volkoff and Buchdahl limits, and quasiblack holes}",
      journal = {\prd},
         year = 2013,
        month = oct,
       volume = {88},
       number = {8},
          eid = {084023},
        pages = {084023},
          doi = {10.1103/PhysRevD.88.084023},
archivePrefix = {arXiv},
       eprint = {1309.4470},
 primaryClass = {gr-qc},
       adsurl = {https://ui.adsabs.harvard.edu/abs/2013PhRvD..88h4023A}
}

@article{Will_2014,
   title={The Confrontation between General Relativity and Experiment},
   volume={17},
   ISSN={1433-8351},
   url={http://dx.doi.org/10.12942/lrr-2014-4},
   DOI={10.12942/lrr-2014-4},
   number={1},
   journal={Living Reviews in Relativity},
   publisher={Springer Science and Business Media LLC},
   author={Will, Clifford M.},
   year={2014},
   month=jun }

@article{Akiyama_2019,
doi = {10.3847/2041-8213/ab0ec7},
url = {https://dx.doi.org/10.3847/2041-8213/ab0ec7},
year = {2019},
month = {apr},
publisher = {The American Astronomical Society},
volume = {875},
number = {1},
pages = {L1},
author = {The Event Horizon Telescope Collaboration },
journal = {The Astrophysical Journal Letters}
}

@article{shapiro:1964prl,
  title = {Fourth Test of General Relativity},
  author = {Shapiro, Irwin I.},
  journal = {Phys. Rev. Lett.},
  volume = {13},
  issue = {26},
  pages = {789--791},
  numpages = {0},
  year = {1964},
  month = {Dec},
  publisher = {American Physical Society},
  doi = {10.1103/PhysRevLett.13.789},
  url = {https://link.aps.org/doi/10.1103/PhysRevLett.13.789}
}

@ARTICLE{taylor:1982,
       author = {{Taylor}, J.~H. and {Weisberg}, J.~M.},
        title = "{A new test of general relativity - Gravitational radiation and the binary pulsar PSR 1913+16}",
      journal = {\apj},
         year = 1982,
        month = feb,
       volume = {253},
        pages = {908-920},
          doi = {10.1086/159690},
       adsurl = {https://ui.adsabs.harvard.edu/abs/1982ApJ...253..908T}
}

@article{Ligo,
  title = {Observation of Gravitational Waves from a Binary Black Hole Merger},
  author = {Abbott},
  collaboration = {LIGO Scientific Collaboration and Virgo Collaboration},
  journal = {Phys. Rev. Lett.},
  volume = {116},
  issue = {6},
  pages = {061102},
  numpages = {16},
  year = {2016},
  month = {Feb},
  publisher = {American Physical Society},
  doi = {10.1103/PhysRevLett.116.061102},
  url = {https://link.aps.org/doi/10.1103/PhysRevLett.116.061102}
}

@article{Bertotti2003,
  author       = {Bertotti, B. and Iess, L. and Tortora, P.},
  title        = {A test of general relativity using radio links with the Cassini spacecraft},
  journal      = {Nature},
  volume       = {425},
  number       = {6956},
  pages        = {374--376},
  year         = {2003},
  doi          = {10.1038/nature01997},
  publisher    = {Nature Publishing Group},
  note         = {\url{https://doi.org/10.1038/nature01997}}
}

@article{chauvineau:2024prd,
  title = {Exact post-Newtonian parameters about spherical stars in Bergmann-Wagoner-Nordtvedt scalar-tensor gravity},
  author = {Chauvineau, Bertrand},
  journal = {Phys. Rev. D},
  volume = {110},
  issue = {6},
  pages = {064060},
  numpages = {7},
  year = {2024},
  month = {Sep},
  publisher = {American Physical Society},
  doi = {10.1103/PhysRevD.110.064060},
  url = {https://link.aps.org/doi/10.1103/PhysRevD.110.064060}
}

@article{nordvert:1968,
  title = {Equivalence Principle for Massive Bodies. II. Theory},
  author = {Nordtvedt, Kenneth},
  journal = {Phys. Rev.},
  volume = {169},
  issue = {5},
  pages = {1017--1025},
  numpages = {0},
  year = {1968},
  month = {May},
  publisher = {American Physical Society},
  doi = {10.1103/PhysRev.169.1017},
  url = {https://link.aps.org/doi/10.1103/PhysRev.169.1017}
}

@article{wagoner:1970,
  title = {Scalar-Tensor Theory and Gravitational Waves},
  author = {Wagoner, Robert V.},
  journal = {Phys. Rev. D},
  volume = {1},
  issue = {12},
  pages = {3209--3216},
  numpages = {0},
  year = {1970},
  month = {Jun},
  publisher = {American Physical Society},
  doi = {10.1103/PhysRevD.1.3209},
  url = {https://link.aps.org/doi/10.1103/PhysRevD.1.3209}
}

@article{Bergmann1968,
  author    = {Peter G. Bergmann},
  title     = {Comments on the scalar-tensor theory},
  journal   = {International Journal of Theoretical Physics},
  year      = {1968},
  volume    = {1},
  number    = {1},
  pages     = {25--36},
  month     = may,
  doi       = {10.1007/BF00668828},
  url       = {https://doi.org/10.1007/BF00668828},
  issn      = {1572-9575}
}

@misc{thomascode,
  title = {Codes and scripts to compute all the figures},
  author = {Chehab, Thomas and Minazzoli, Olivier},
  year = 2025,
  doi = {10.5281/zenodo.20606137}
}

@article{PREM,
title = {Preliminary reference Earth model},
journal = {Physics of the Earth and Planetary Interiors},
volume = {25},
number = {4},
pages = {297-356},
year = {1981},
issn = {0031-9201},
doi = {https://doi.org/10.1016/0031-9201(81)90046-7},
url = {https://www.sciencedirect.com/science/article/pii/0031920181900467},
author = {Adam M. Dziewonski and Don L. Anderson},
}

@article{model_S,
author = {J. Christensen-Dalsgaard  and W. Däppen  and S. V. Ajukov  and E. R. Anderson  and H. M. Antia  and S. Basu  and V. A. Baturin  and G. Berthomieu  and B. Chaboyer  and S. M. Chitre  and A. N. Cox  and P. Demarque  and J. Donatowicz  and W. A. Dziembowski  and M. Gabriel  and D. O. Gough  and D. B. Guenther  and J. A. Guzik  and J. W. Harvey  and F. Hill  and G. Houdek  and C. A. Iglesias  and A. G. Kosovichev  and J. W. Leibacher  and P. Morel  and C. R. Proffitt  and J. Provost  and J. Reiter  and E. J. Rhodes  and F. J. Rogers  and I. W. Roxburgh  and M. J. Thompson  and R. K. Ulrich },
title = {The Current State of Solar Modeling},
journal = {Science},
volume = {272},
number = {5266},
pages = {1286-1292},
year = {1996},
doi = {10.1126/science.272.5266.1286},
URL = {https://www.science.org/doi/abs/10.1126/science.272.5266.1286},
eprint = {https://www.science.org/doi/pdf/10.1126/science.272.5266.1286},}

@article{Damour_1996,
   title={Tensor-scalar gravity and binary-pulsar experiments},
   volume={54},
   ISSN={1089-4918},
   url={http://dx.doi.org/10.1103/PhysRevD.54.1474},
   DOI={10.1103/physrevd.54.1474},
   number={2},
   journal={Physical Review D},
   publisher={American Physical Society (APS)},
   author={Damour, Thibault and Esposito-Farèse, Gilles},
   year={1996},
   month=jul, pages={1474–1491} }

@ARTICLE{Minazzoli_CQG_2012,
       author = {{Minazzoli}, Olivier},
        title = "{2PN/RM gauge invariance in Brans-Dicke-like scalar-tensor theories}",
      journal = {Classical and Quantum Gravity},
         year = 2012,
        month = dec,
       volume = {29},
       number = {23},
          eid = {237002},
        pages = {237002},
          doi = {10.1088/0264-9381/29/23/237002},
archivePrefix = {arXiv},
       eprint = {1210.3073},
 primaryClass = {gr-qc},
       adsurl = {https://ui.adsabs.harvard.edu/abs/2012CQGra..29w7002M}
}

@article{Potekhin2004:AA,
	author = {{Potekhin, A. Y.} and {Fantina, A. F.} and {Chamel, N.} and {Pearson, J. M.} and {Goriely, S.}},
	title = {Analytical representations of unified equations of state for neutron-star matter},
	DOI= "10.1051/0004-6361/201321697",
	url= "https://doi.org/10.1051/0004-6361/201321697",
	journal = {AA},
	year = 2013,
	volume = 560,
	pages = "A48",
	month = "",
}

@ARTICLE{Douchin2001:AA,
       author = {{Douchin}, F. and {Haensel}, P.},
        title = "{A unified equation of state of dense matter and neutron star structure}",
      journal = {AA},
         year = 2001,
        month = dec,
       volume = {380},
        pages = {151-167},
          doi = {10.1051/0004-6361:20011402},
archivePrefix = {arXiv},
       eprint = {astro-ph/0111092},
 primaryClass = {astro-ph},
       adsurl = {https://ui.adsabs.harvard.edu/abs/2001A&A...380..151D}
}

@misc{deniscode,
  title = {Codes and scripts with the piece wise equation of state},
  howpublished = {\url{https://github.com/denisArruga/ER_hotspot}},
  author = {Arruga, Denis},
  year = 2022,
}
\setcounter{section}{0}
\renewcommand\thesection{\Alph{section}}

\section{Validating our integration model}
\label{app:validate}

\begin{figure}[htpb]
    \centering
		\includegraphics[scale=0.4]{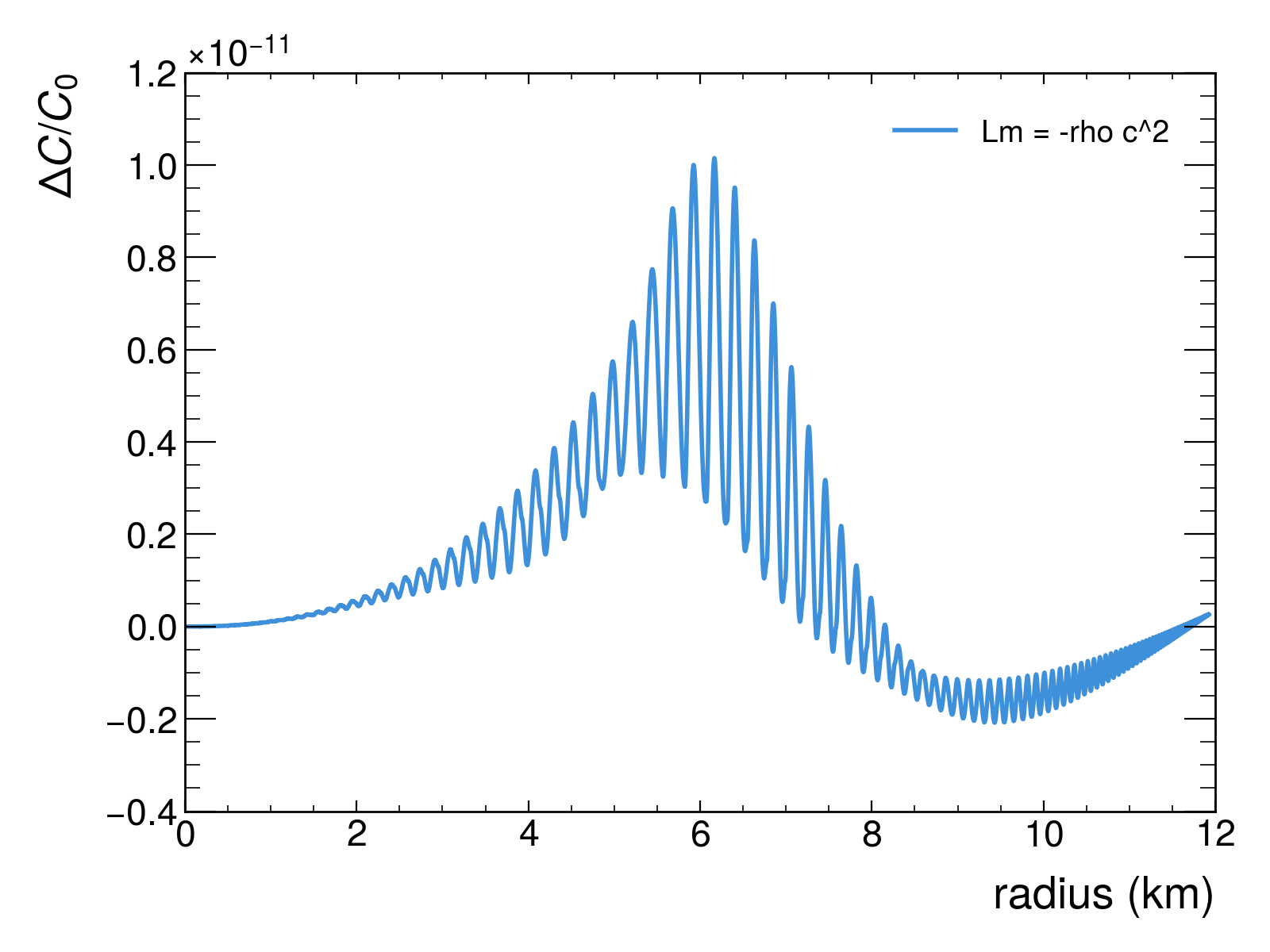}
\caption{Plot of the relative variation of the constant given by $\frac{C_i - C_0}{C_0}$ throughout the numerical integration, versus the radius in Entangled Relativity, for a central density of $100$ MeV/fm$^3$.}\label{plot:first_int_ER}
\end{figure}

Checking the consistency of our integration can be achieved through the use of a first integral. Indeed, by direct integration of Eq. (\ref{eq:dpdr_BD}), one can obtain a conserved quantity given by
\be \label{eq:constant_BD}
C = 5 \ln{\left(\frac{c^2}{k^{3/5}}+ P^{2/5}\right)} + \ln{a},
\ee
where $C$ is a constant. As a reminder, $k = 1.475 \times 10^{-3}$ (fm$^3$/MeV)$^{2/3}$. $P$ is the pressure and $a \equiv g_{tt}$. This quantity can be computed from the initial values and monitored throughout the integration process. Verifying that this constant remains indeed constant allows us to check the consistency of our integration. We find that our integration, both for Brans–Dicke and Entangled Relativity\footnote{In Entangled Relativity, Eq. (\ref{eq:constant_BD}) is modified by an additive term $\ln{\phi}$.} theories, leads to relative variations of the constant of order $10^{-11}$ or below\footnote{The variation increase with increasing density. A relative variation of the order of $10^{-11}$ is the maximum for both theories.}, as illustrated in Fig. \ref{plot:first_int_ER}.

\end{document}